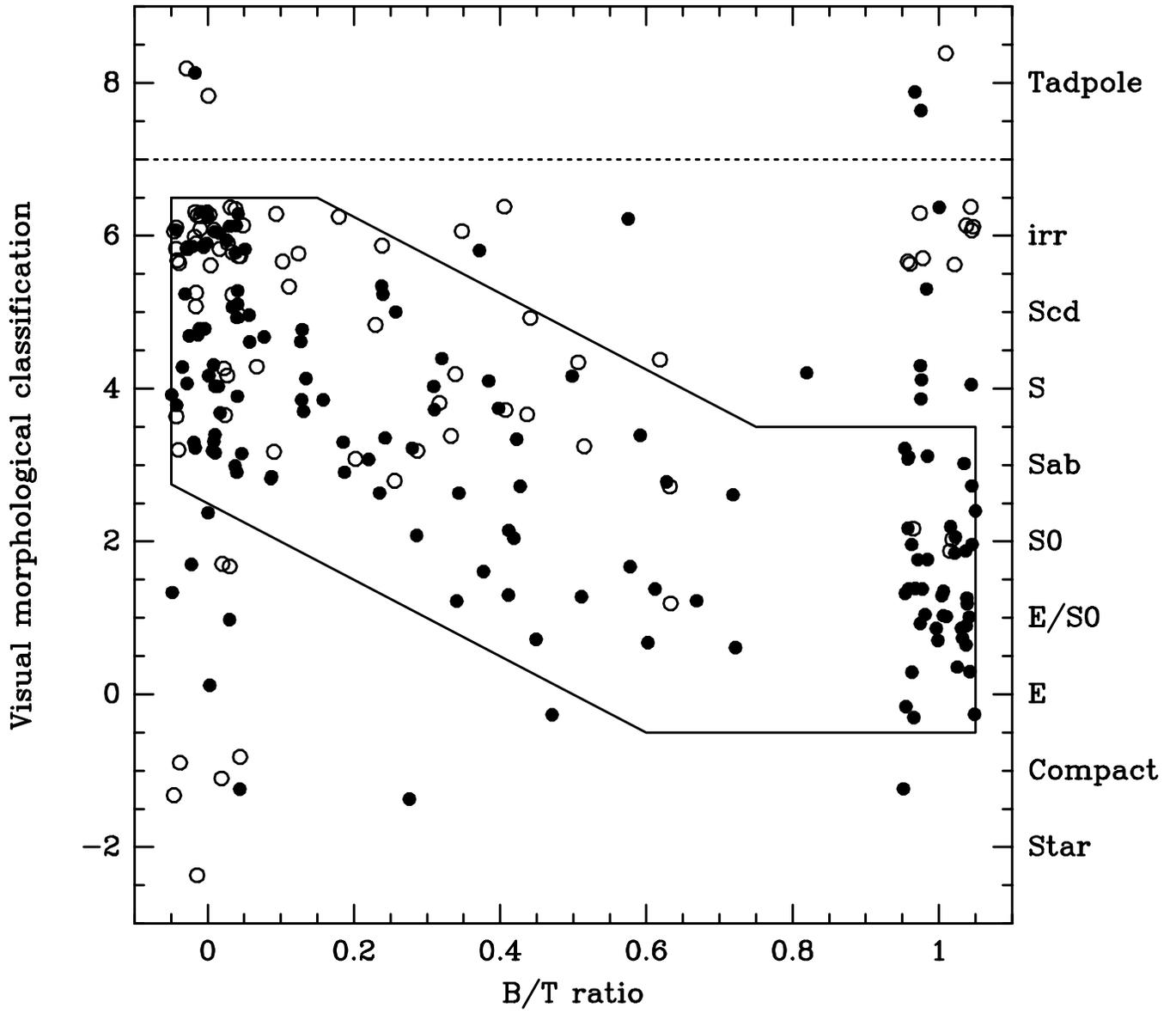

Lilly et al Fig 1.

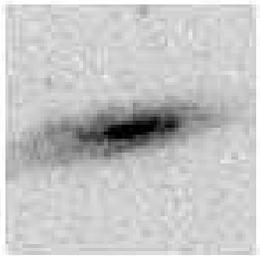 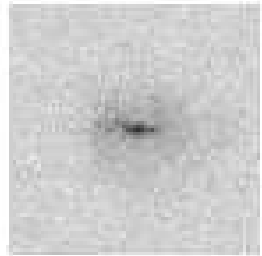 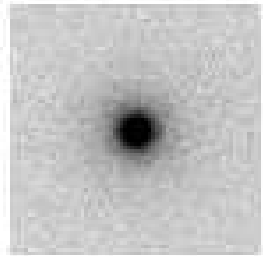 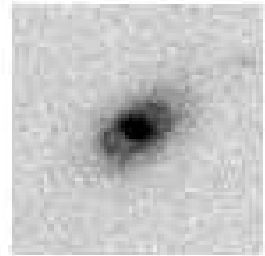
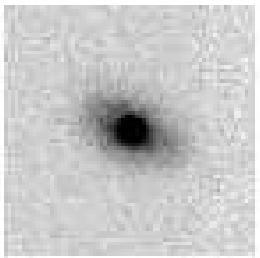 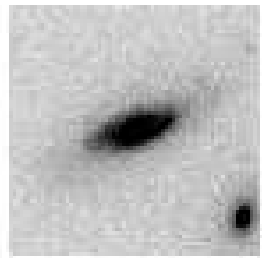 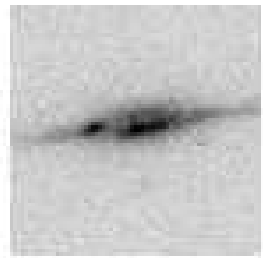 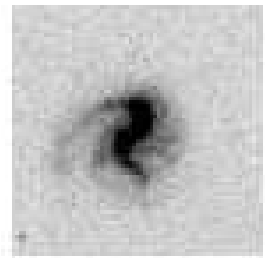
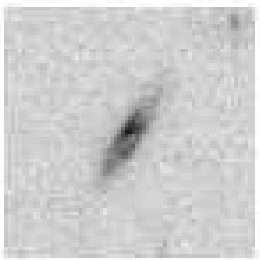 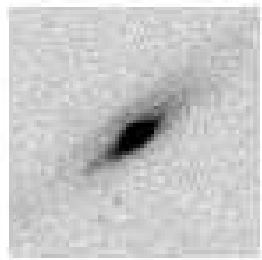 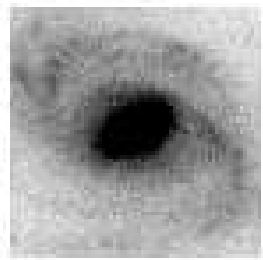 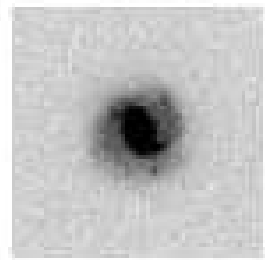
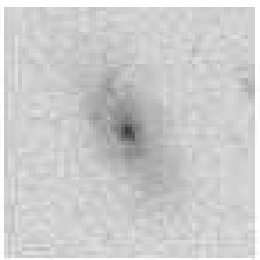 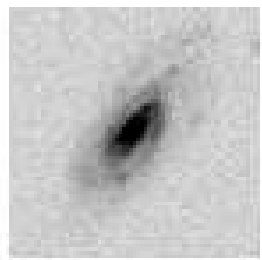 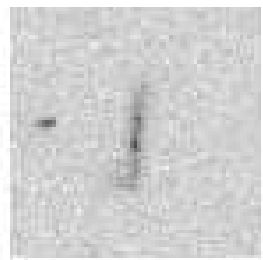

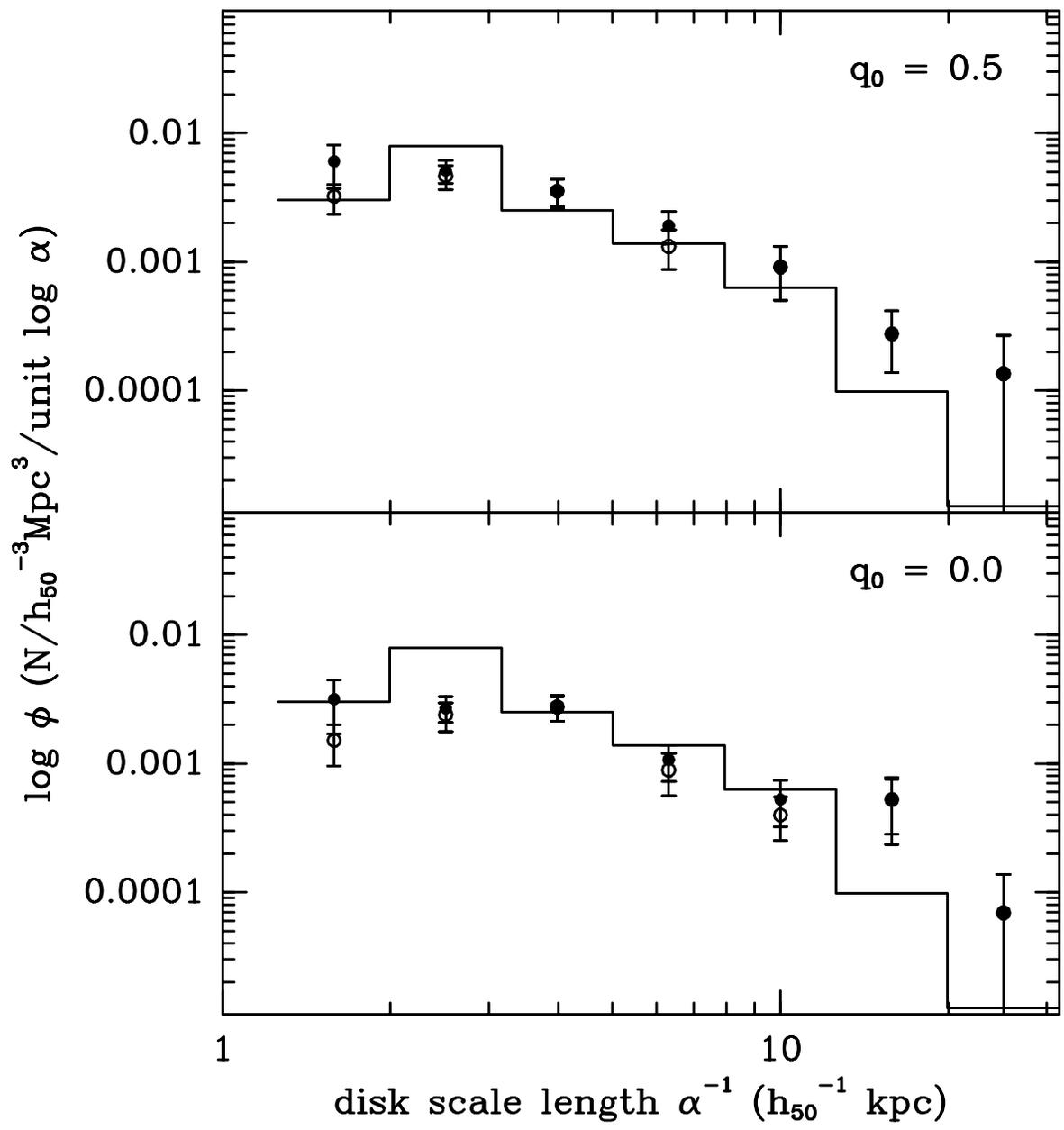

Lilly et al Fig 2

**Hubble Space Telescope imaging of the CFRS and LDSS redshift surveys II:**

**Structural parameters and the evolution of disk galaxies to *z* ~ 1.**

Simon Lilly[1,2], David Schade[3], Richard Ellis[2], Olivier Le Fevre[4,5], Jarle Brinchmann[2], Roberto Abraham[6], Laurence Tresse[2], Francois Hammer[4], David Crampton[3], Matthew Colless[7], Karl Glazebrook[8], Gabriela Mallen-Ornelas[1], Thomas Broadhurst[9]

[1]Department of Astronomy, University of Toronto, Canada
[2]Institute of Astronomy, University of Cambridge, United Kingdom
[3]Dominion Astrophysical Observatory, Victoria, Canada
[4]Observatoire de Paris, Meudon, France
[5]Laboratoire d'Astronomie Spatiale, Marseilles, France
[6]Royal Greenwich Observatory, Cambridge, UK
[7]Australian National University, Canberra, Australia
[8]Anglo-Australian Observatory, Epping, Australia
[9]University of California, Berkeley, U.S.A.

**Abstract**

Several aspects of the evolution of star-forming galaxies are studied using measures of the 2-dimensional surface brightness profiles of a sample of 341 faint objects selected from the CFRS and LDSS redshift surveys that have been observed with the Hubble Space Telescope. The size function of disk scale lengths in disk-dominated galaxies (i.e. bulge to total ratios, $B/T \leq 0.5$) is found to stay roughly constant to $z \sim 1$, at least for those larger disks with exponential scale lengths $\alpha^{-1} > 3.2\ h_{50}^{-1}$ kpc, where the sample is most complete and where the disk and bulge decompositions are most reliable. This result, which is strengthened by inclusion of the local de Jong et al (1996) size function, suggests that the scale lengths of typical disks can not have grown substantially with cosmic epoch since $z \sim 1$, unless a corresponding number of large disks have been destroyed through merging. In addition to a roughly constant number density, the galaxies with large disks, $\alpha^{-1} \geq 4\ h_{50}^{-1}$ kpc, have, as a set, properties consistent with the idea that they are similar galaxies observed at different cosmic epochs. However, on average, they show higher B-band disk surface brightnesses, bluer overall (U-V) colors, higher [OII] 3727 equivalent widths and less regular morphologies at high redshift than at low redshift, suggesting an increase in the star-formation rate by a factor of about 3 to $z \sim 0.7$. This is consistent with the expectations of recent models for the evolution of the disk of the Milky Way and similar galaxies. The evolution of the large disk galaxies with scale lengths $\alpha^{-1} \geq 4\ h_{50}^{-1}$ kpc, is probably not sufficient to account for the evolution of the overall luminosity function of galaxies over the interval $0 < z < 1$, especially if $\Omega \sim 1$. Analysis of the half-light radii of all the galaxies in the sample and construction of the bivariate size-luminosity function suggests that larger changes in the galaxy population are due to smaller galaxies, those with half-light radii around 5 $h_{50}^{-1}$ kpc (i.e. disk scale lengths of 3 $h_{50}^{-1}$ kpc or less).

## 1: Introduction

Considerable progress has recently been made in observing the cosmic evolution of the population of galaxies. The systematic measurement of redshifts of large numbers of faint galaxies in the redshift range $0 < z < 1.3$, as in the CFRS (Lilly et al 1995a and references therein), LDSS (Glazebrook et al 1995, Ellis et al 1996) and Hawaii Deep Survey programs (Cowie et al 1996), have yielded a broadly consistent description of changes in the galaxy luminosity function over the last half to two-thirds of the history of the Universe (see e.g. Lilly et al 1995c). At higher redshifts, the isolation of $z > 2.3$ galaxies through the ``Lyman-break'' color selection technique (Steidel et al 1996) has enabled an estimate of the evolution of the integrated comoving luminosity density in the Universe (which likely tracks the global star-formation rate) to be constructed over the entire range $0 < z < 4$ (Lilly et al 1996, Madau et al 1996, Connolly et al 1997).

It is clear that the largest changes in the luminosity function are associated with galaxies that have blue colors (Lilly et al 1995c, Heyl et al 1997) and/or high [OII] 3727 equivalent widths (Ellis et al 1996). Nevertheless, the physical processes responsible for this evolution to $z \sim 1$ have not been convincingly identified. This is partly due to the limitations of the basic photometric and redshift data from the redshift survey programs, and also to fundamental difficulties that are encountered in attempting to associate particular galaxies at different epochs. Almost all observationally accessible quantities (such as luminosities, colors, spectra, morphologies, masses and comoving space densities) may plausibly change as an individual galaxy evolves. This difficulty is further compounded by the fact that available samples of galaxies necessarily only sample a restricted part of the galaxy population, implying selection criteria whose effects must be carefully considered in the context of the large dispersion in properties exhibited by even well-defined sub-samples.

This paper is the second of a series of papers that combine high quality morphological information obtained with the Hubble Space Telescope (HST) with redshift information from the existing deep CFRS and LDSS redshift surveys. In an earlier paper in this series, Brinchmann et al (1997, hereafter Paper 1) analyzed the morphological classifications of these galaxies, using both visual classifications and machine-based algorithms that can be calibrated for the effects of the shifting rest-frame bandpass of the HST observations. The analysis in Paper 1 showed that the main changes in the galaxy population were associated with galaxies with late-type (i.e. "irregular/peculiar") morphologies — strengthening the conclusions that had been drawn from the classification of large samples of faint galaxies without individual redshift information in the HST Medium Deep Survey (Glazebrook et al 1995, Driver et al 1995) and in the Hubble Deep Field (Abraham et al 1996).

In this paper, we adopt an approach to galaxy morphology that is based primarily on the sizes and surface brightnesses of the galaxies derived from modeled fits to the 2-dimensional light distributions of the galaxies. This is an extension of earlier analyses of HST and CFHT imaging of smaller subsets of the CFRS sample (Schade et al 1995, 1996a). This paper concentrates on the properties of the star-forming galaxies, and in particular on those with large disks, while a companion paper (Schade et al 1997, hereafter Paper 3) is concerned with the properties of the spheroidal population. The sample selection and the surface brightness fitting process are briefly reviewed in Section 2 (both have been covered in more detail in earlier articles) along with some important methodological considerations regarding selection effects that are associated with this approach.

Three aspects of the size measurements of star-forming galaxies are then discussed in this paper. First, in Section 3, we construct the metric size function for galactic disks over the redshift range $0.2 < z < 1.0$ as a basic description of the galaxy population. At least for the larger disks, those with exponential scale lengths $\alpha^{-1} \geq 3.2\ h_{50}^{-1}$ kpc, the size function is found to be roughly constant with look-back time and consistent with local estimates (de Jong 1996b).

We therefore then look in Section 4 at the average properties of the "large disk" galaxies, i.e. those with disk scale length $\alpha^{-1} \geq 4\ h_{50}^{-1}$ kpc, making the implicit assumption that these form an identifiable class of galaxy whose evolution can be studied in isolation from the rest of the galaxy population. There are several attractions in concentrating on the largest galaxies for this detailed study. First and foremost, for these large galaxies it is likely that our sample is more or less "complete" to high redshifts (i.e. all galaxies of normal surface brightness will be included). Thus, following the changes in the average properties of these galaxies with redshift should allow us to track the evolution of this particular class of galaxy. Second, these galaxies have large angular sizes, $\alpha^{-1} \geq 0.5"$ for all $z \leq 1$ and the quantitative analysis of their light profiles and morphologies is relatively straightforward, particularly with HST. Finally, the conclusions regarding the evolution of this class of galaxy can be tested against the ``fossil-record'' of similar galaxies studied locally, including our own Milky Way Galaxy. In Section 5 of the paper, the results from the preceding two sections are discussed in the context of our expectations of how galaxies similar to the Milky Way have evolved.

Although a number of independent evolutionary effects are seen in these large galaxies, they are insufficient to account for the changes seen in the galaxian luminosity function or in the overall luminosity density, particularly if $\Omega \sim 1$. So, in the final section of the paper, Section 6, we analyze the sizes (half-light radii) of all the galaxies in the sample in order to identify the sizes of the galaxies producing the largest changes in the bivariate size-luminosity function.

The paper is summarized in Section 7. Throughout the paper, we adopt a Hubble Constant of $H_0 = 50 h_{50}$ kms$^{-1}$Mpc$^{-1}$ and, except where indicated, take $q_0 = 0.5$, although it should be noted that many of the quantities derived from the data are largely independent of the choice of $q_0$.

## 2: Structural parameters - methodology

### 2.1 HST observations of the CFRS-LDSS sample

The HST observations and the sample have been described in detail in Paper 1 and only a brief summary is provided here. The analysis is based on HST F814W images of 25 WFPC2 fields that contain 251 objects from the CFRS (Le Fevre et al 1995, Lilly et al 1995b, Hammer et al 1995) and 90 from the LDSS (Ellis et al 1996) spectroscopic surveys. The bulk of the imaging data has come from our own Cycle 4 and 5 imaging data, supplemented by archival images of the Groth strip (Groth et al 1994) that bisects the CFRS 1417+52 field. Of the total sample of 341 sources, 30 are stars, 4 are quasars and 35 are galaxies with unknown or insecure redshifts. The remaining 272 are galaxies with measured redshifts in the interval $0.01 < z < 1.3$. The galaxies span a wide range in luminosity (see Figure 1 of Paper 1).

Both the CFRS and LDSS surveys are nominally magnitude limited, based on deep isophotal photometry that should closely approximate total photometry (see e.g. Lilly et al 1995a, and Paper 1 for a discussion). The CFRS subsample is *I*-band selected and has a median $z \sim 0.60$ whereas the LDSS subsample is *B*-band selected and has a median $z \sim 0.38$. In this paper, where we attempt to analyze "complete" samples of galaxies, the extra depth and smaller fraction of galaxies without measured redshifts of the CFRS sample (9% versus 21%) is often important, and so some of the statistical analyses have been restricted to the CFRS galaxies alone, although the LDSS objects are included in the discussion of the properties of individual galaxies in Section 4.

### 2.2 2-dimensional fitting procedure

The 2-dimensional fitting procedure has been described in detail elsewhere (Schade et al 1995, 1996, Paper 3), and only a summary is given here. As a first step, each image has subtracted from it a version of itself rotated by 180º, producing an "asymmetric residual" image. This is then set to zero below a threshold of +2σ and the resulting image thus represents any positive components of the galaxies that are significantly asymmetric. Next, a "symmetrized" galaxy is obtained by subtracting this asymmetric residual image from the original galaxy image. A fitting radius is then defined from the growth curve and all pixels within that radius are fit with a two-dimensional galaxy surface brightness model that has been convolved with the point spread function.

The full galaxy model is in principle defined by 10 parameters: a center ($x,y$), an exponential disk component characterized by central surface brightness $\mu_0$, scale length $\alpha^{-1}$, axial ratio $b/a$, and position angle $\theta$, plus a de Vaucouleurs $r^{1/4}$ spheroid component also defined by a surface brightness, effective radius, axial ratio and position angle. In practice, the two position angles are constrained to be the same, and the integrated brightness of the model within the sampling radius is also constrained to be that observed in aperture photometry, so there are eight free parameters. In addition, 6 parameter "pure bulge" and "pure disk" models (with 5 free parameters) were also fit to each galaxy. The goodness of fit of different models within the optimization scheme was assessed using a straightforward $\chi^2$ statistic. Choice between the 5-parameter and 8-parameter models was based on inspection of a final residual image obtained by subtracting the chosen "best fit" from the original image. It should be noted that the number of pixels located within the sampling radius is very much larger than the number of parameters in the fit.

The appropriateness of the final model was quantified using the $R_A$ and $R_S$ parameters (defined by Schade et al 1995) which indicate the fraction of the brightness of the galaxy that can not be represented by the symmetric two-dimensional two-component models. The quantitative values from the fits are clearly more reliable, and presumably more meaningful for galaxies with small residuals. In the following sections we will distinguish between galaxies according to the sum of $R_A$ and $R_S$ parameter, designated by $R_{A+S}$. For reference, two-thirds of the large disk sample analyzed in this paper have $R_{A+S} \leq 0.10$. This is a more stringent criterion than that adopted in Schade et al (1995) who used $R_{A+S} \leq 0.15$.

The parameters of the separate disk and spheroid components are then used to define a bulge to total light ratio, *B/T*. Figure 1 shows the *B/T* values derived from the fits against the visual classification from Paper 1 of all objects in the sample with the exception of stars and quasars. For clarity, galaxies the points are displaced from their actual locations by a small random offset. Solid symbols represent fits with $R_{A+S} \leq 0.1$ and open symbols represent fits with $R_{A+S} > 0.10$. Overall there is a high degree of consistency between the *B/T* derived from the fits and the visual classification. The irregularly shaped region of the diagram indicates the area where the fits and visiual classification are "consistent". It is noticeable that most of the galaxies which lie outside of this area have poor residuals. The objects in the top right of Fig. 1 which are fitted as spheroids but classified as irregular are the "blue nucleated galaxies" identified by Schade et al 1995. The objects in the bottom left are very compact galaxies for which an exponential light profile is as reasonable as any other.

*2.3 Methodological considerations*

Until recently, rather little use has been made of structural parameters of high redshift galaxies as an evolutionary diagnostic, and there are some important methodological considerations that are also relevant for the analysis of kinematic data (see e.g. Rix et al 1997, Vogt et al 1996, 1997, Guzman et al 1997).

First, there is the issue of whether the sizes of star-forming galaxies will change with time. Even in isolated spiral galaxies, infall patterns and the varying efficiency of star-formation with radius may lead to changes in the apparent disk scale length $\alpha^{-1}$ (see Section 5 and the references therein). In extreme hierarchical models in which merging and morphological transformation are common (Kauffmann et al 1993, Baugh et al 1996) disks that are present at early epochs may be completely destroyed to be replaced by new disks later, so the "size" of the disk in a particular "galaxy" may change dramatically. Thus, any assumption of a constant scale length for disks must be viewed with some caution. Our own approach to this question will be through construction of the size-function for galactic disks. For a stable population of isolated disks, any systematic change in disk scale length should produce a change in the size function, though it should be noted that a constant size function could also be produced by growing disks if the number density decreases, perhaps through the destruction of disks in mergers.

Second, the use of size as an "identifier" of galaxies and thus of surface brightness as a diagnostic of luminosity evolution may introduce selection biases when applied, as here, to samples of galaxies that were originally selected by apparent magnitude. Regardless of any surface brightness selection biases present in the original sample (believed to be small in the case of the CFRS, see Lilly et al 1995a) and in the identification of disk components in the HST images, the apparent magnitude cut imposes a surface brightness limit that will vary with the size of the galaxy — small galaxies of low surface brightness will be excluded from the sample simply because their low integrated luminosities fall below the selection cut-off for the sample. The problem is, of course, exacerbated by the wide dispersion in surface brightness seen in galaxies of the same scale-size. The changing effect of such a luminosity/surface brightness selection function with redshift could be mistaken for evolutionary changes in the population. In this paper, two approaches are taken to address this issue. First, the size function is studied to see if there is evidence for a significant number of galaxies being "lost" at high redshift. Secondly, the redshift range over which each individual galaxy would be visible within the original magnitude-limited sample is analyzed. If this extends throughout the redshift range of interest then the effects of this bias are likely to be small.

Against these somewhat negative considerations, there are of course many attractions of using surface brightness as a quantitative indicator of galactic evolution. First, the average surface brightness of galaxies changes only slowly with luminosity and size, so that uncertainties in the composition of the sample due to the variation of the derived sizes (or luminosities) of galaxies with $q_0$, should not produce significant uncertainties in the average surface brightnesses of the sample. In addition, the observational determination of surface brightness is formally independent of $q_0$, as are color, line strength and morphological classification. Furthermore, it is found that these other diagnostics of star-formation activity correlate only weakly with surface brightness (due to the large dispersion in the latter) and thus the distribution of these should be largely unaffected by any selection biases that may be operating in surface brightness.

## 3. The size function of galactic disks

The size function for galactic disks, i.e. the number of disks per unit comoving volume per unit logarithmic interval in scale length, $\phi(\alpha^{-1})$, has been computed using those CFRS galaxies that are disk dominated (i.e. have bulge to total ratios, $B/T < 0.5$). Construction of the size function utilizes the $V_{max}$ formalism (following the procedures outlined by Lilly et al 1995c, Schade et al 1996) For each galaxy with measured redshift within some redshift interval of interest, $z_1 \leq z \leq z_2$, the minimum and maximum redshifts, $z_{min}$ and $z_{max}$, between which the object would satisfy the photometric selection criteria of the original redshift surveys are calculated, deriving $k$-corrections from the observed $(V-I)_{AB}$ colors. The accessible volume, $V_{max}$, is then the volume between $\max(z_1, z_{min})$ and $\min(z_2, z_{max})$. It is important to note that, especially for the larger galaxies of most interest in this paper, the accessible volume is usually bounded by $z_1$ and $z_2$ rather than by $z_{min}$ and $z_{max}$, and thus $V_{max}$ is largely independent of the parameter defining the sample. Thus the size-function should be insensitive to problems associated with a non-uniform density distribution, allowing the use of the simple $V_{max}$ approach.

The size function is computed as the sum over all galaxies in this redshift range:

$$\phi(\alpha^{-1}) \, d(\log \alpha) = \Sigma \, 1/V_{max}$$

Following Lilly et al (1995c), $1\sigma$ uncertainties in the size function have been estimated using a boot-strap approach. No attempt has been made to account for the increase in the uncertainties arising from the clustering of galaxies within the small field of view of the WFPC2. Where no objects were observed in the sample, an upper limit has been derived as the number density representing one galaxy within the average $V_{max}$ for that redshift bin. The few objects for which a secure redshift was not obtained in the original spectroscopic surveys have been treated by (a) initially ignoring them and (b) then including them at their photometrically estimated redshifts (Crampton et al 1995).

It is important to note that the size functions $\phi(\alpha^{-1})$ calculated here are based on *all* the galaxies with sufficient luminosities to appear above the original *I*-band magnitude limit of the CFRS — in effect the bivariate $\phi(L,\alpha^{-1})$ is integrated down to a limiting luminosity that is a function of redshift. Changes in the size function at small sizes should thus be treated with caution since these may arise because of the change in limiting luminosity.

In Fig. 2, the size function for the CFRS sample between $0.5 < z < 1.0$ is shown, for two values of $q_0$. The open symbols are based on the disk galaxies with secure redshifts, the solid symbols show the effects of including the spectroscopically unidentified large disk galaxies at their photometric redshifts. Qualitatively, there is little dependence on cosmology because the effects of the different sizes and volumes tend to cancel out.

The high redshift $\phi(\alpha^{-1})$ is compared with the local size function estimated by de Jong et al (1996b). This is based on a similar profile-fitting analysis of a sample of 86 disk-dominated spiral galaxies selected from the UGC catalogue to have major axis $\geq 2$ arcmin, axial ratio $\geq 0.625$ and galactic latitude $\geq 25^o$ (de Jong 1996a). As with most measures of the galaxy population at high and low redshift, there are significant differences in the method of construction of this local size function compared with that presented here and the generation of a local size function that is more directly reliable to the high redshift one should be considered a priority for the future. Nevertheless the consistency of the luminosity function constructed from the de Jong sample with the Kirschner et al (1983) luminosity function (see de Jong 1996a) is reassuring. It should be noted that the fitting functions in the de Jong (1996b) are not identical to those used here (with an exponential bulge rather than a de Vaucouleurs $r^{1/4}$ bulge) . However, this difference in fitting functions is unlikely to affect the derived disk parameters for the disk-dominated galaxies with large scale lengths studied here (de Jong 1996b).

In order to investigate changes in the size function with redshift, Fig. 3 shows the $q_0=0.5$ CFRS-based size function $\phi(\alpha^{-1})$ for three redshift bins $0.2 < z < 0.5$, $0.5 < z < 0.75$ and $0.75 < z < 1.0$. In each case, a double power-law consistent with the de Jong (1996b) local size function has been fit to the high redshift $\phi(\alpha^{-1})$ between $3.2$ kpc $< \alpha^{-1} < 32$ kpc, allowing both the size and density normalizations to independently vary. The range of acceptable fits (i.e. those yielding the minimum $\chi^2 +1$) that is obtained at each redshift is shown in Fig. 4, normalized to the best-fit to the de Jong size function. It should be noted that the high redshift size functions are based on essentially volume limited samples of galaxies, whereas the local de Jong (1998b) size function is based on a sample that is selected on apparent size. Although these should be equivalent, the local size functions is relatively well defined at large sizes (because these galaxies are selected from a larger volume). Thus, in fitting the local size-function to the high redshift data, a wide combination of characteristic (size,density) combinations are allowed along a diagonal locus in Fig. 4. In particular, it should be noted that a scenario with a higher density of smaller galaxies in the past (as might be expected in a strongly merger-driven hierarchical picture) would be permitted by the $0.2 < z < 0.5$ and $0.75 < z < 1.0$ size functions (in the latter there are actually no galaxies with $\alpha^{-1} > 12$ kpc), although there are several galaxies with $\alpha^{-1} > 12\ h_{50}^{-1}$ kpc in the $0.5 < z < 0.75$.

In Fig. 5, the results of fitting the $z = 0$ size function to the high redshift data by varying only the size (i.e. keeping the comoving density fixed) and by varying only the density (keeping the sizes fixed) are shown as a function of redshift. There is little evidence for a significant reduction in the sizes of galactic disks at a constant comoving density or, equivalently, in the number of disk galaxies at a constant size. Formal fits to the points in Fig. 5 as $n \propto (1+z)^\nu$ yield $\nu = -0.02 \pm 0.44$ for the CFRS points at $z \geq 0.2$. As is often the case, the uncertainty is formally reduced if the redshift baseline is extended to include the larger de Jong (1996b) sample at low redshift, with the associated concerns about systematic uncertainties. In this case the formal fit yields $\nu = 0.13 \pm 0.17$

The constancy of the size function with redshift for disks with scale length $\alpha^{-1} > 3.2\ h_{50}^{-1}$ kpc is an interesting result that will be discussed further in Section 5. This result makes it plausible that (a) disk scale lengths are roughly constant with epoch since $z \sim 1$ and that (b) the CFRS sample of "large disks" (with $\alpha^{-1} \geq 4\ h_{50}^{-1}$ kpc) probably contains most such galaxies at all $z \leq 1$. It is thus a reasonable exercise to study the properties of the $\alpha^{-1} \geq 4\ h_{50}^{-1}$ kpc sample in isolation, in order to deduce evolutionary changes in typical large, disk-dominated, spiral galaxies similar to the Milky Way. This is the subject of the next section of the paper. As an aside, the choice of $\alpha^{-1} \geq 4\ h_{50}^{-1}$ kpc as the size cutoff was made before the final computation of the size function using the de Jong et al (1996b) binning scheme, and the extra 30% in size in any case gives an extra 0.5 magnitude margin with regard to the surface brightness selection effects (see Section 4.3.1 and Fig. 8 below).

# 4 Large disk galaxies with $\alpha^{-1} > 4\,h_{50}^{-1}$ kpc

## 4.1 The sample of large disk galaxies

The 42 galaxies in the CFRS/LDSS sample with measured $z$ whose 2-dimensional surface brightness profile fits have $B/T \leq 0.5$ and $\alpha^{-1} \geq 4\,h_{50}^{-1}$ kpc are listed in Table 1 and shown in a montage of 7 arcsec "postage stamps" in Plates 1, 2 and 3 for the three redshift bins $0.2 < z < 0.5$, $0.5 < z < 0.75$, $0.75 < z < 1.0$ respectively. In each plate, the galaxies are arranged in the order (left to right, top to bottom) that they appear in Table 1  In the analysis below, we also consider an additional 4 CFRS galaxies for which no redshift was securely measured and which would qualify for the $\alpha^{-1} \geq 4\,h_{50}^{-1}$ kpc sample if they lie at high redshift. This they almost certainly do based on their color-estimate redshifts which are in the 0.7 to 0.8 range (Crampton et al 1995). In computing the average properties of the sample galaxies with redshift we exclude these galaxies, but indicate on the associated figures the rest-frame properties that these galaxies would have at a range of redshifts.

## 4.2 Consistency checks

In this section, several properties of these large disk galaxies are examined, primarily to support the case that these galaxies form a long-lived and isolatable class of galaxies observed at different redshifts. The galaxies are also found to be broadly similar in these properties to the de Jong (1996a) sample at low redshifts.

*4.2.1 Axial ratios and inclination effects*

Two dimensional disks randomly oriented in space should have a uniform distribution in disk axial ratio *b/a* between 0 and 1 (although intrinsic asymmetries will cause an avoidance of *b/a* = 1 and finite disk thickness will likewise cause an avoidance of *b/a* = 0). Fig 6 shows the distribution of axial ratio, *b/a*, for the galaxies in the three redshift bins. The distribution is reasonably uniform at each redshift, confirming that these components with exponential profiles are indeed two-dimensional disks (c.f. Im et al 1995).

Assuming the relationship between the surface brightness and disk axial ratio as:

$$\mu_{0,obs} = \mu_{0,face\ on} - 2.5\ C \log (a/b)$$

transparent disks will have $C = 1$ and optically thick disks will have $C < 1$, with the value depending on a number of factors, including the geometries of the stars and dust and the relative importance of scattering and absorption. The optical thickness of disks at low redshift is the source of much debate (see e.g. Davies et al 1993 and Simien et al 1993 and references therein), and so we adopt here an empirical approach. The three panels in Fig. 7 show the variation of central surface brightness (with cosmological effects removed - see Section 4.4.1 below) with axial ratio *b/a* in the three redshift bins $0.2 < z < 0.5$, $0.5 < z < 0.75$ and $0.75 < z < 1.0$.. In each redshift bin, no significant correlation is seen between the inclination and the observed central surface brightness of the disk. Formal fits for *C* using a straightforward least squares algorithm yield $C = 0.04 \pm 0.33$, $C = 0.12 \pm 0.35$ and $C = 0.0 \pm 0.36$ for the three bins (in order of increasing *z*). Given this and the uncertainty at low redshift, no inclination correction has been applied to the present data. It should be noted that application of an inclination correction would act to decrease the implied surface brightnesses of the disks by an average of 1.1×*C* for a sample that was uniformly distributed in *b/a*.

*4.2.2 Bulge to total light ratios*

Another consistency check comes from the distribution of bulge to total light ratios (*B/T*). This is shown for the three redshift bins in Fig. 8. On the left side, the B/T values directly observed in the F814W images are shown. However, the F814W passband samples longer rest-wavelengths at lower redshifts and so any color differences between bulge and disk will cause the bulge to be more prominent at lower redshifts. A small reduction (based upon assuming the spectral energy distribution of the bulge is that of an elliptical and that of the disk is the spectral energy distribution of the CWW Scd galaxy) has been applied to the observed *B/T* ratios to estimate the value that would be observed in the rest-frame *B*-band. This reduction is typically 20% of the *B/T* at the low redshifts and decreases with redshift. The resulting ``corrected'' distributions are shown in the right hand side of Fig. 8, together with that of the de Jong et al (1996a) local sample which is also measured in rest B-band. Except at the highest redshift $z > 0.75$, where the number of galaxies is small, the distributions are evidently similar, again consistent with the idea that these galaxies represent a homogeneous class of galaxy seen at different redshifts. Of course, differential evolution between bulge and disk could change *B/T*, although it is found that the surface brightness evolution for both is similar, as expected if passive evolution plays a dominant role.

**4.3 Potential incompleteness effects**

Since the motivation of this section is to compare the average properties of galaxies selected by a particular size criterion at different redshifts and thus to trace the evolution of typical members of that class of galaxy, a major concern is whether the sample is biased against some particular members of that class at any redshift. Two potential biases are immediately obvious.

*4.3.1 Surface brightness effects*

There are several ways that surface brightness selection effects can enter into the sample. First, there are obvious observational difficulties of detecting extreme low surface brightness galaxies, either in the initial ground-based imaging (see e.g. Lilly et al 1995a) or in detecting very low surface brightness disks in the current HST imaging.

Surface brightness biases can also arise in a more subtle way through the use of magnitude limited samples. As noted in Section 2.3, the parent CFRS sample is selected to have an isophotal (close to total) magnitude $I_{AB} \geq 22.5$. A consequence of this is that for galaxies of a given *size*, those with low *surface brightnesses* will be excluded below a certain surface brightness threshold which will be a function of size (decreasing to larger sizes) and redshift (increasing to higher redshifts).

Aside from noting the constancy of the size function in Section 3, the potential of the luminosity-related effect to give trouble has been examined using the $z_{max}$ values computed in Section 3. If the $z_{max}$ of a particular galaxy extends well into and, ideally, throughout, the next higher redshift bin then it implies that this galaxy would still have been detected in the next higher redshift bin, even assuming no luminosity evolution, and thus that any comparison in average properties, for galaxies of this particular type, should be meaningful between these two bins. If on the other hand, the $z_{max}$ does not extend into the next higher redshift bin, then these objects would be missed from that bin, unless a luminosity increase had occurred to brighten them back into the sample. In this case the higher redshift bin would potentially be biased with respect to the lower redshift one. Of the 11 large disk galaxies seen in the CFRS sample at $0.2 < z < 0.5$, 7 have $z_{max} > 0.75$ and would be visible throughout the next higher redshift bin, 3 have $0.5 < z_{max} < 0.75$ and would only be visible over a significantly reduced volume, and one has $z_{max} = 0.43$ and would not be visible at all in the higher redshift sample, unless by boosted evolutionary effects. Of the 15 large disks observed in the CFRS sample at $0.5 < z < 0.75$, 10 have $z_{max} \geq 1$ and would be visible throughout the $0.75 < z < 1.0$ range, 2 have $0.75 < z_{max} <$ 1.0, and 3 have $z_{max} < 0.75$. Clearly, it is possible (in the absence of evolution) to lose about 20-30% of the sample between adjacent redshift bins.

The mean surface brightness in the lower redshift $0.2 < z < 0.5$ sample that is obtained by *eliminating* all those galaxies that would *not* be visible throughout the next higher redshift bin, $0.5 < z < 0.75$, is 0.20 magnitudes higher than when these galaxies are included. This is in principle the maximum effect that the bias could have on the average surface brightness between these redshift bins.

Finally, we can examine the de Jong (1996a) sample of large disk galaxies and compute the mean surface brightness, assuming no evolution, that is obtained as a function of redshift by eliminating those individual galaxies that would fall below the CFRS magnitude limit. The mean surface brightness is constant to $z \sim 0.4$ (since no galaxies are eliminated) and then increases by 0.1 mag to $z = 0.6$ and by 0.4 mag at $z = 1.0$ (see Fig 9).

*4.3.2   Incomplete redshift determinations*

In the CFRS sample studied here, there are three galaxies for which spectroscopic observations did not yield a redshift and which would have large disks $\alpha^{-1} > 4h_{50}^{-1}$ kpc if they have large redshifts, as they most likely do. There is also one large disk galaxy with an insecure redshift (i.e. Confidence Class 1, see Le Fevre et al 1995) at $z = 0.88$, The photometrically-estimated redshifts of the three failures (see Crampton et al 1995), which have a nominal uncertainty of $\sigma_z = 0.2$, are 0.70, 0.74 and 0.83. Thus, these four galaxies are likely to lie in one or the other of the two higher redshift bins in our analysis. In the following analyses of the properties of large galaxies, we have constructed the locus of their properties that is obtained by placing them at all plausible redshifts within the redshift range being considered. This plausible range is set by the ($V$-$I$)$_{AB}$ color which should not imply an intrinsic spectral energy distribution that is redder than an unevolved elliptical galaxy.

## 4.4 Tracers of star-formation: the average properties of large disk galaxies

In the previous sections it has been argued that the CFRS sample of disk-dominated galaxies with scale-lengths $\alpha^{-1} > 4h_{50}^{-1}$ kpc forms a long-lived isolatable sample of galaxies that is consistent with representing, at least statistically, similar objects seen at different *z*. In this section, we will examine those properties that are most relevant for tracing the history of star-formation in the galaxies. These are (a) the rest-frame *B*-band surface brightness; (b) the overall rest-frame (*U-V*) color; (c) the [OII] 3727 equivalent width and (d) the morphology of the galaxy as classified by eye. In each case, we will differentiate between the statistically complete CFRS sample (including the objects without measured redshifts) and the larger sample that includes the LDSS galaxies with measured redshifts. We will also differentiate between including all galaxies and confining attention to those "best behaved" ones, i.e. with residuals $R_{A+S} \leq 0.1$ and morphological classifications (Paper 1) of "spiral".

### 4.4.1 Surface brightness

The observed central surface brightness of the disks obtained from the 2-dimensional fits in the observed F814W band, and uncorrected for any inclination effects (see Section 3.1), has been converted to a rest-frame $B_{AB}$ by applying the cosmological dimming term and a *k*-correction color term. Following the notation of Lilly et al (1995c):

$$\mu_0(B_{AB}) = \mu_0(\text{F814W}_{AB}) - 2.5 \log (1+z)^3 + (B\text{-F814W}_z)_{AB}$$

In the absence of color information on the disk components in isolation, the color *k*-correction term $(B\text{--}F814W_z)_{AB}$ is determined from matching the observed integrated (3 arcsec aperture) ground-based colors, $(V\text{-}I)_{AB}$ for the CFRS and $(B\text{-}R)$ for the LDSS samples, using the procedure described in Lilly et al (1995c). At high redshifts, this term is small since F814W is redshifted down towards the rest-frame *B*-band (they coincide at z = 0.83). At lower redshifts, this term is larger and, since the overall galaxy may be redder than the disk component, this procedure may lead to an underestimate of the *B*-band surface brightness. We estimate that this could amount to as much as 0.25 mag for the worst-case galaxy with $B/T = 0.5$ at $z = 0.3$, but only 0.12 mag at *z*- 0.6.

The resulting central surface brightnesses in the rest-frame B-band are shown in Fig. 9 along with the comparison sample of de Jong et al (1996), which has a central surface brightness close to the canonical Freeman (1970) value of $\mu_{AB}(B) = 21.6$. The mean surface brightnesses are shown in Table 2. In each case, the means, dispersions and formal errors in the mean have been computed for the CFRS sample alone, and with the LDSS objects added, and also by separately considering all objects and then only those "well-behaved" objects which are morphologically classified as spirals (Class ≤ 5, paper 1) and have $R_{A+S} \leq 0.1$. In no case are the conclusions driven by the inclusion of either the *B*-selected LDSS objects or the less well-behaved objects.

The dispersion in central surface brightness within the population is large (as expected) but roughly constant with redshift except at the largest redshifts $z > 0.75$ where the biases against low surface brightness galaxies are strongest. There is clearly a trend towards higher central surface brightnesses at higher redshifts and the average central surface brightnesses at $0.2 < z < 0.5$ are in the range $21.3 < \mu_{AB}(B) < 21.7$ (depending on the sample) with a formal statistical uncertainties for each sample of around 0.30, and at $0.5 < z < 0.75$ it is $20.65 < \mu_{AB}(B) < 20.85$ with a statistical uncertainty of 0.25. These latter are considerably higher than the Freeman (1970) value of $\mu_{AB}(B) = 21.6$ and the surface brightnesses of the de Jong sample, $21.75 \pm 0.13$.

We estimate that the net observed effect is thus about 0.8 ± 0.3 magnitudes to $z = 0.67$, or an increase proportional to $(1+z)^{1.4\pm0.5}$.

The dashed lines in Fig. 9 explore the effect of the surface brightness biases discussed above. First, the irregular dot-dash curve shows the mean surface brightness computed from the de Jong (1996a) sample of $\alpha^{-1} \geq 4\ h_{50}^{-1}$ kpc disks but excluding those galaxies that, at each redshift, would fall below the CFRS luminosity selection criteria. The short dashed line shows the minimum central surface brightness required by an $\alpha^{-1} = 4\ h_{50}^{-1}$ kpc disk galaxy with $B/T = 0$ if it is to satisfy the original CFRS selection criteria. The long dashed line in Fig. 9 shows the surface brightness corresponding to $\mu_{AB}(814) \sim 24.5$, which is coincidentally both the limiting surface brightness detection limit of the HST F814W images ($1\sigma$ per 0.01 arcsec$^2$ WFPC2 pixel) and the observed central surface brightness limit (after the effects of seeing) of the original CFRS sample Lilly et al 1995a). This is unlikely to have a larger effect than the luminosity-selection bias discussed above. As noted above, as much as 0.2 magnitudes of the 0.5 magnitude change between $0.2 < z < 0.5$ and $0.5 < z < 0.75$ could be produced by this effect, and we believe our estimate of 0.8 magnitudes of evolution to $z \sim 0.7$ could conceivably be overestimated by as much as 0.3 magnitudes.

The effect of the galaxies without measured redshifts on the observed increase is unlikely to be large. The dotted lines in Fig. 9 show the derived surface brightnesses loci for the three CFRS galaxies. Clearly, placing these galaxies at any particular redshift along these loci will not affect the general conclusion of modest surface brightness evolution in the sample.

### 4.4.2  *Overall $(U-V)_{0,AB}$ color*

In the absence of HST images at shorter wavelengths, the colors of the disk components can unfortunately not yet be studied in isolation. Therefore, the (*V-I*) and (*B-R*) integrated colors of the galaxies obtained from ground-based photometry have been converted to an estimate of the rest-frame $(U\text{-}V)_{AB}$ color using the color/SED-matching method described above and in more detail by Lilly et al (1995c). These are shown in Fig. 10. The average colors of the large-disk galaxies at high redshift are about 0.5 magnitudes bluer at $0.5 < z < 0.75$ than at $0.2 < z < 0.5$ although the latter are in fact on average redder than the local de Jong sample (see Fig 13) so the color change may be smaller. At a purely empirical level, Fig. 10 suggests that at least some individual galaxies have crossed the red-blue color divide (shown as a dashed line) that was used by Lilly et al (1995c) to define the differential evolution of the galaxy luminosity function. As in the previous Section, inclusion of the galaxies without measured redshifts at any redshift would not alter this conclusion, since the loci of these galaxies with varying assumed redshift mimics the general trend seen in the galaxies with redshifts.

Furthermore, the dependence of surface brightness on color is weak. Representing

$$\mu_{0,AB}(B) = A\ (U\text{-}V)_{0,AB} + constant$$

gives $A = 0.25 \pm 0.3$, $A = 0.25 \pm 0.5$ and $A = 0.22 \pm 0.4$ in the three redshift bins (and $A = 0.34 \pm 0.3$ in the de Jong sample), with only marginally significant correlation coefficients. Thus the color distribution is not likely to be driven by any surface brightness biases. The effect of applying the CFRS luminosity cut to the de Jong (1996) sample, as a function of redshift, is shown as the irregular dot-dash curve. Observationally, the ground-based $(V\text{-}I)_{AB}$ and $(B\text{-}R)_{AB}$ colors are also completely independent of the HST $\mu$(F814W) surface photometry.

*4.4.3    [OII] 3727 equivalent width*

Additional evidence for increased levels of star-formation activity comes from the [OII] 3727 line. Fig. 11 shows the distribution in rest-frame equivalent widths of [OII]3727 in the sample, taken from Hammer et al (1997) for the CFRS and unpublished estimates for the LDSS.

The median equivalent width probably increases with redshift, but the increase is modest, probably about 50% over the redshift range studies, especially when it is remembered that the objects without measured redshifts will likely have weak or absent emission lines. A 50% increase in equivalent width, coupled with a 0.5-0.8 magnitude increase in *B*-band surface brightness (Section 4.4.1) and a roughly 0.3 magnitude decrease in (U-V) (Section 4.4.2) suggests that the total [OII] 3727 luminosities of these galaxies are likely to be 2.5 to 3.5 times higher at $z = 0.7$ than locally.

*4.4.4  Morphological classifications*

Finally, and at a more descriptive level, Fig. 12 shows the distributions of visual morphological classifications for these galaxies using the system defined in Paper 1. The distribution of morphologies shifts to later types at higher redshifts. The median shifts from Class 4 (``mid-spiral") at $0.2 < z < 0.5$ to Class 5 (``Sdm") at $0.5 < z < 0.5$, and 50% of the $\alpha^{-1} \geq 4$ kpc disks at $0.75 < z < 1.0$ were classified as having "Irregular/Peculiar"' morphologies. Some of this shift to later morphologies will be due to straightforward wavelength-dependencies of the morphology as discussed above in the context of the *B/T* ratio. In Paper 1, up to 25% of the galaxies in the Frei et al sample (1996) that were classified as Spirals on simulated F814W images at low redshift, would be classified as Irr at $z \sim 1$  However, the F814W bandpass is well-matched to the rest-frame *B*-band at $z = 0.9$, so the high incidence of Irregular-like morphologies most likely reflects a real increase in the irregularity of these galaxies, plausibly arising from more vigorous star-formation activity, or from an increased incidence of interactions and mergers between galaxies (Le Fevre et al, in preparation).

## 4.5 Summary: evidence for modestly increased star-formation in large disks at earlier epochs

The best indicators of star-formation rates, such as the Hα luminosity, are unfortunately unavailable at present for high redshift galaxies. Nevertheless, the changes in mean properties identified above give a consistent picture of increased star-formation at high redshift.

Fig 13 shows a color-luminosity plot derived from the Bruzual and Charlot (1993) GISSEL library of solar metallicity stellar population models. Stellar population models with exponentially declining star-formation rates and Salpeter initial mass functions (with $x = 1.35$) are shown. The luminosities of the models are normalized at 13 Gyr and converted to a relative surface brightness (assuming constant physical area). The colors have been reddened by E(*U-V*) = 0.2 mag of extinction and by the addition of a red bulge component that contributes 20% of the light at *V*. The diagram makes the interesting point that, as long as the star-formation peaked at earlier epochs, the *B*-band surface brightness change is not *strongly* dependent on the star-formation history (because in this case there is a substantial passive component to the evolution) but is maximized for a model in which the star-formation rate has an exponential decay time of about 3 Gyr. Fig 13 also illustrates why the observed surface brightness evolution of the disk and spheroid components of different galaxy components can be similar (c.f. Schade et al 1996b) resulting in a roughly constant observed *B/T* ratio (e.g. Fig 8). Only models with an exponential decay time longer than the age of the Universe have an qualitatively different luminosity evolution with the luminosity monotonically increasing with cosmic time.

The choice of a particular model for the star-formation history thus relies more on the colors than on the luminosity change. This is unfortunate because the colors are sensitive to reddening and contamination from a spheroid component. Nevertheless, the available data points from de Jong (1996a) and from Table 2 broadly match the track in Fig 13 expected for a model with an e-fold decline of around 5 Gyr and do not match the track expected for constant star-formation. The $0.2 < z < 0.5$ point is around $2\sigma$ from its expected position in such a scenario, but it should be remembered that the corrections used to derive both the rest frame $(U-V)_{AB,0}$ and $\mu_{B,0}$ of the galaxies from the $(V-I)_{AB}$ and the $\mu_{814}$ measurements are largest for this lowest redshift bin. It should also be noted that the surface brightness of the $z = 0.63$ point may be overestimated by about 0.2 mag and that at $z = 0.87$ by as much as 0.4 mag, due to the surface brightness selection effects discussed above.

A star-formation history with an e-fold decline of around 5 Gyr, as suggested from Fig 13, would imply an increase in the star-formation rate by a factor of about 4 back to $z \sim 0.75$ (a look-back time of around 7.5 Gyr). This is consistent with the estimated increase in [OII] 3727 luminosity derived in Section 4.4.4, since this would be expected to track the star-formation rate, if the nature of the star-formation (e.g. the initial mass function) remained unchanged (see Kennicutt 1992, Gallagher et al 1989).

## 5. Discussion: the evolution of large disk galaxies

In the previous two sections it has been argued that large disk-dominated galaxies (a) have a size function that does not change strongly to $z \sim 1$ and (b) have surface brightnesses, colors, [OII] 3727 equivalent widths and visual morphologies that are all consistent with a modestly elevated star-formation rate, by a factor of about 3, at $z \sim 0.75$. In this section, these results are discussed in the context of previous results on high redshift galaxies and in the context of our understanding of the evolution of the disk of the Milky Way.

### 5.1 Consistency with earlier morphological and structural studies

There have by now been several attempts to estimate the increase in average surface brightness enhancements of disk galaxies at high redshifts, although comparisons of these in the literature have not generally taken into account the sample selection criteria. Schade et al (1995, 1996) estimated the mean surface brightness samples of CFRS disk galaxies at $z > 0.5$ from HST imaging of 15 galaxies and from high resolution CFHT imaging of 107 galaxies respectively. In both cases, a change in surface brightness of around 1.2 magnitudes (Schade et al 1995) and 1.6 magnitudes (Schade et al 1996) relative to the Freeman value was found at a median redshift of about 0.75. At first glance, this is 1.5 to 2 times larger than the effect derived in the present analysis, but it is important to appreciate that these earlier values were computed for a straight magnitude (i.e. luminosity) limited sample, since a major motivation was to understand changes in the CFRS luminosity function. The analysis in this paper has been based on a size-selected sample, in order to track the evolution of a single class of galaxy. If the straight luminosity selection includes a large number of small high surface brightness galaxies, as is indeed the case (see Fig. 8 of Schade et al 1996, Guzman et al 1997 and Section 6 below), then we would *expect* to see a smaller effect in the size-selected sample (c.f. the discussion by Vogt et al 1997).

The sample selection in the Forbes et al (1996) and Vogt et al (1997) samples is not particularly well-defined, but is likely selected primarily on size and visual appearance. The estimation of surface brightnesses was also somewhat simpler (being based on simple major axis profiles rather than a full 2-dimensional decomposition including the effects of the point spread function). Both these studies derive surface brightness changes of 0.6 mag at a median redshift of $z = 0.5$, very similar to that derived here.

The brightness evolution discussed here is also consistent with the conclusion of Paper 1, that the number of spiral galaxies in the sample is consistent with their having brightened by one magnitude or less to $z = 1$, when it is remembered that the comoving density of disks at the highest redshifts may have dropped by a modest fraction (Section 3) and that some of the large disks at high redshift studied here are classified as "Irregular/Peculiar" in Paper 1, and thus would not appear in the spiral histogram (see Fig 11 of Paper 1 and Fig 12 here).

## 5.2 Comparison to models for the evolution of the Milky Way Galaxy

The scale length of the stellar disk in the Milky Way Galaxy is still uncertain but is likely to be in the $3 \leq \alpha^{-1} \leq 5$ kpc range (e.g. Lewis & Freeman 1989, Ortiz & Lepine 1993). Thus the Milky Way would likely satisfy the $\alpha^{-1} \geq 4\ h_{50}^{-1}$ kpc criteria of the present sample, especially if $h_{50} > 1$, and the wealth of detailed data on our own Galactic disk allows the study of the "fossil record" of the evolutionary changes directly observed at high redshift.

Higher star-formation rates in the Galactic disk in the past have been indicated since the classic analysis of the age-metallicity relation in the solar neighborhood by Twarog (1980) which indicated an elevated star-formation rate by a factor of 2-3 when the disk was 1/2 to 1/3 of its present age. The more comprehensive theoretical models for the evolution of the disk as a function of radius depend on several poorly understood phenomena, including the nature of infall onto the disk as $f(t,R)$ and the physical parameters that control the star-formation rate (SFR) at different galactic radii.

A detailed examination of different model predictions is beyond the scope of this paper and is, in any case, probably not yet warranted by the limited observational data at high redshifts. However, we note that the results derived above, concerning the size function and the changes in average star-formation rates, compare rather well with the "toy model" preferred by Frantzos & Aubert (1995) in their comparison of the predictions of a series of generic models with a "minimal set" of present-epoch measurements on the Galactic disk, including the age-metallicity relation in the solar neighborhood, the radial distributions of stars and gas and the star-formation rate, and the radial dependence of oxygen and iron abundances. In particular, this model (their Fig 12) predicts, for a decrease of a factor of 2 in age (as would be appropriate for $z = 0.6$ for early formation in an $\Omega = 1$ cosmology), a decrease in the scale length of the radial dependence of star-formation of only 18% (from 3.75 kpc to 3.2 kpc) and an increase in the SFR at a radius $R = 7$ kpc of a factor 2.3. A factor of 3 change in age increases these to 25% and a factor of 2.5 respectively.

Similarly, the more recent model by Bouwens et al (1997) for the population of spiral disks, which is based on the de Jong (1996b) size function, yields a population of $\alpha^{-1} \geq 4\ h_{50}^{-1}$ kpc disks at $z \sim 1$ which is reduced in number relative to today, by between 11% ($\Omega = 0$) and 29% ($\Omega = 1$).

**5.3 The contribution to the overall evolution of the galactic population to $z = 1$**

It was argued above that large disk galaxies as a class, exhibit a surface brightness increase of at most $\Delta\mu_B \sim 1.1 \times z$. This corresponds to $L_B \propto (1+z)^{1.4}$ at $z \sim 0.7$. Coupled with a size function that is constant, the comoving luminosity density must increase rather more slowly than was estimated for the Universe as a whole by Lilly et al (1996), see also Connolly et al (1997). It should be noted that the shortfall is larger for the case of $\Omega = 1$, where the global $B$-band luminosity density was estimated increase as $(1+z)^{2.7\pm0.5}$, than for $\Omega = 0$, for which the increase is $(1+z)^{2.2\pm0.5}$. Nevertheless, it is likely that the relatively mild evolution of the large disk galaxies as described above is not the main contributor to the evolution of the galaxian luminosity function and the overall luminosity density of the Universe back to $z \sim 1$, and that more rapid evolution of some other population is implied. A similar conclusion is reached in Paper 1 from study of the morphological classifications.

Accordingly, in the next Section we examine the sizes of all of the galaxies in the current sample as a function of their location in the luminosity-color plane and construct the bivariate size-luminosity function $\phi(L, r_{0.5})$.

## 6. The sizes of the most strongly evolving component of the galaxy population

In the previous two sections of the paper we have used morphological decompositions and size measurements to extract a sub-sample of galaxies which is arguably complete at all redshifts of interest and which can therefore be used to trace the evolution of a particular class of galaxy. In this final section of the paper, we take an orthogonal approach, looking at the half-light sizes of all of the galaxies. Many of the smaller galaxies have quite irregular morphologies (Paper 1) and the physical meaningfulness of the multi-parameter two-dimensional fits for these galaxies is less clear than for the larger and more regular ones studied in previous sections. The two-dimensional fits can however usefully be used to derive half-enclosed-light radii for all the galaxies. The motivation for using the modeled-fits is (a) to avoid systematic isophotal effects related to estimating total brightnesses, (b) to eliminate strongly asymmetric components in an effort to trace the "underlying" light profile, (c) to consider disk and spheroid components equally and to avoid ambiguities between the two for small galaxies, and (d) to mitigate seeing effects, although these should not be large for the HST data. It should be noted that the radii are estimated from the two-dimensional fits as if along the major axis — in effect, non-circularity is assumed to arise from the inclination of a flattened circularly symmetric disk and the galaxies are de-inclined before the half-light radii are computed.

We first look at the half-light sizes of galaxies selected from a particular location on the evolving luminosity function in order to see what size of galaxy is producing the largest changes in the galaxy population as identified by the bivariate color-luminosity function (Lilly et al 1995c, Heyl et al 1996). Fig 14 shows the luminosity-size plane, split in panels as functions of redshift (increasing to the right) and rest-frame color (top and bottom). Galaxies shown as asterisks are the $\alpha^{-1} > 4\ h_{50}^{-1}$ disks analyzed above, while those represented by triangles satisfy the selection criteria of the sample of "compact" galaxies studied by Phillips et al (1997) and Guzman et al (1997) . Open circles represent galaxies with estimated redshifts. The heavy dashed line represents the approximate luminosity limit corresponding to the I-band magnitude limit of the CFRS (see Lilly et al 1995c).

The diagram illustrates how the area of size-luminosity space occupied by the large disks studied here, and kinematically by Vogt et al (1996, 1997), do not show strong evolutionary effects — the number of objects appears roughly constant. In contrast the biggest change as we go to the higher redshifts is in the number of smaller galaxies, those with half-light radii around $h_{50}^{-1}$ kpc (i.e. between 2 and 8 $h_{50}^{-1}$ kpc). This is especially apparent in the parts of the figures representing the bluer galaxies. The more compact ones would satisfy the Guzman/Phillips selection criteria, where the evolutionary changes inferred from kinematic data are larger (Guzman et al 1997) than for the large disks (Vogt et al 1996,1997).

To see this more directly, the bivariate size-luminosity function $\phi(r_{0.5},L)$ is shown in Figure 15 and tabulated in Table 3. This has been computed using the same formalism as in Section 3 using the half-light radii defined as above, for both $q_0 = 0.5$ and $q_0 = 0.0$. Because of the limited numbers of objects, there are only two bins for redshift, $0.2 < z < 0.5$ and $0.5 < z < 1.0$, three logarithmic bins in size (with $\Delta \log r_{0.5} = 0.5$) centered on 0.89, 2.8 and 8.9 $h_{50}^{-1}$ kpc and four luminosity bins centered on $M_{AB}(B) = -19.5, -20.5, -21.5, -22.5$ (although the lowest luminosity bin is omitted at the higher redshift since it contains virtually no objects). In each bin in Table 3, the maximum, nominal and minimum values of $\phi(r_{0.5},L)$ are shown (the range covering 2/3 of the Monte Carlo realizations) vertically side-by-side for the two redshift ranges, together with the net change, and estimated uncertainty, of the change from low redshift to high. These are shown on Figure 15.

Although the $\phi(r_{0.5},L)$ function is noisy, two features are apparent. First, the part of the diagram associated with the largest galaxies (the "ridge" along the back of Figure 15) shows only small changes (Table 3). Indeed, none of the $r_{0.5} > 5\ h_{50}^{-1}$ kpc bins (corresponding to disk scale lengths $> 3\ h_{50}^{-1}$ kpc) at any luminosity show a significant excess at high redshift, although all are elevated by about 0.2 dex (Table 3), the typical uncertainty. The biggest change occurs in the large increase (almost an order of magnitude increase for $q_0 = 0.5$) in the number of luminous small galaxies (i.e. with $-20 > M_{AB}(B) > -21$ and $1.5 < r_{0.5} < 5\ h_{50}^{-1}$) that fill in at $z > 0.5$ the "re-entrant" seen at $z < 0.5$.

Thus, while the evolutionary changes identified in the large disks in the first part of the paper must undoubtedly contribute to the changes in the luminosity function, the smaller galaxies appear to have a larger role, particularly due to the appearance of a large number of relatively small galaxies with relatively high luminosities. The nature of these galaxies, and their present day descendants is not well understood at this point, although they may well simply be the small disk galaxies seen today at $M_B \leq -19$. The fact that large disk galaxies with $\alpha^{-1} > 3.2\ h_{50}^{-1}$ kpc are present to $z \sim 1$ in roughly the numbers seen today suggests that these smaller galaxies are *not* the antecedents of the large disk galaxies seen today.

It should however be noted that the indications for strongly differential galactic evolution are reduced as $q_0$ is lowered. As the assumed value of $q_0$ is lowered, the reduction in implied number densities and increase in the sizes and luminosities allow the larger galaxies to play a larger role in the evolution. This is seen in Table 3 where the "excess" values are rather more uniform across size-luminosity space in the $q_0 = 0.0$ case than for $q_0 = 0.5$. This behavior with $q_0$ is also seen in the $\phi(L,z)$ luminosity function presented by Lilly et al (1995c). In the low $q_0$ case, the high redshift luminosity function allows and/or requires substantial luminosity evolution in the luminous systems, whereas the higher densities and lower luminosities produced by the high $q_0$ case produces more of a "piling up" of galaxies at moderate luminosities around L*, which would be a signature of differential effects

This $q_0$ dependence on the interpretation is a different manifestation of the oft-remarked fact that it is easier to match the galaxy number count and redshift $N(m,z)$ data with "conventional" models with pure luminosity evolution in low density Universes (see e.g. Koo and Kron 1992, Pozetti et al 1996 and references therein). It is also, of course, directly linked to the point about the contribution to the global luminosity density that was made in Section 5.3.

## 7. Summary and conclusions

Quantitative analysis of the two-dimensional light distributions of a sample of 341 objects from the CFRS and LDSS redshift surveys observed with HST has enabled us to reach the following conclusions (largely independent of $q_0$) concerning the evolution of star-forming galaxies over the redshift range $0 < z < 1$.

1. The size function of disk scale lengths for disk-dominated galaxies ($B/T > 0.5$) stays roughly constant to $z \sim 1$, at least for those larger disks where our sample is most complete. This is seen within the sample, $n \propto (1+z)^{\pm 0.5}$, and in comparison with the local size function estimated by de Jong (1996), $n \propto (1+z)^{+0.15 \pm 0.2}$. Assuming that the local sample is compatible with the high redshift data, and that the number of disks has not been reduced through widespread merging, then the scale lengths of typical disks is unlikely to have grown by more than about 25% since $z \sim 1$. A larger degree of growth would require a significantly larger number density of disk-dominated galaxies in the past.

2. As well as having roughly constant number density, the disk-dominated galaxies with large disks, $\alpha^{-1} \geq 4\, h_{50}^{-1}$ kpc, observed over the redshift range $0.2 < z < 1.0$ have, as a set, properties that are consistent with the idea that they are a homogeneous sample of galaxies observed at different cosmic epochs. These properties include the distribution of axial ratios of the disks and the distribution of $B/T$ ratios.

3.  Although there is a large dispersion in their individual properties at all redshifts, the large disk-dominated galaxies with $\alpha^{-1} \geq 4\ h_{50}^{-1}$ kpc show, on average, higher *B*-band disk surface brightnesses, bluer overall (U-V) colors, and slightly higher [OII] 3727 equivalent widths as well as less regular morphologies at high redshift than at low redshift. The surface brightness increases by about 0.8 mag to $z = 0.7$, relative to the Freeman value, with some uncertainty caused by luminosity-related selection biases against lower surface brightness galaxies. This is consistent with previously published estimates, including our own, once differences in the way the results are presented are taken into account. The changes observed in surface brightness, color and line strength are consistent with a model in which the star-formation rate declines with an e-fold time of around 5 Gyr, implying an increase in the star-formation rate by a factor of about 3 to $z \sim 0.7$.

4.  The roughly constant sizes and moderately elevated star-formation rates inferred for the large disk-dominated galaxies in this study are completely consistent with the expectations of recent models for the evolution of the disk of the Milky Way (e.g. Prantzos & Aubert 1995, Bouwens et al 1997).

5.  The evolution of the large disk galaxies with $\alpha^{-1} \geq 4\ h_{50}^{-1}$ kpc, is not large enough to account for all the evolution of the overall luminosity function of galaxies over the interval $0 < z < 1$, especially if $q_0 \sim 0.5$. Analysis of the half-light radii of all the galaxies in the sample suggest that a bigger effect arises from smaller galaxies, with half-light radii of 5 $h_{50}^{-1}$ kpc or less (equivalent to disk scale lengths of 3 $h_{50}^{-1}$ kpc or less). The evidence for differential evolution is weaker if $q_0 \sim 0$.

**Acknowledgments**

We thank the staff of the STScI for their sustained operation of the telescope. SJL's research has been supported by the Raymond and Beverly Sackler Foundation at the Institute of Astronomy in Cambridge and by the National Science and Engineering Council of Canada in Toronto. This collaboration has also been financially supported by the North Atlantic Treaty Organization.

**Figure Captions**

Figure 1   The values of bulge to total ratio, *B/T*, plotted against the visual morphological classification for all the galaxies in the sample (i.e. excluding stars and quasars). Galaxies whose fits had low residuals are represented as solid symbols and those with poorer residuals ($R_{A+S} > 0.1$) as open symbols. For clarity, symbols are displaced by a small random offset. Most of the galaxies lie within the zone of consistency represented by the irregular zone and most of the outliers have in any case poor residuals.

Figure 2   The size-function of disks computed from the CFRS sample of galaxies with *B/T* < 0.5 at $0.5 \leq z \leq 1.0$ for $q_0 = 0.5$ and $q_0 = 0.0$ (points with errorbars), compared with the local size function computed by de Jong et al (1996b) (histogram). Open symbols represent the effect of including only those CFRS galaxies with secure redshifts, while the solid symbols show the effect of adding galaxies that have only photometrically estimated redshifts. The size function is computed from all "accessible" luminosities, and therefore does not include galaxies that do not appear in the original magnitude limited samples on account of their low luminosities and it is therefore unreliable at small sizes. For scale lengths larger than $\alpha^{-1} \sim 3h_{50}^{-1}$ kpc the high redshift size function appears very similar to the local one. The size function at high redshift is not strongly dependent on the assumed value of $q_0$.

Figure 3   As for Fig. 2 except the CFRS sample is split into three redshift ranges, $0.2 < z < 0.5$, $0.5 < z < 0.75$, and $0.75 < z < 1.0$ and plotted for $q_0 = .5$ only. Also shown are the results of fitting a double power-law size-function (derived from the de Jong local size function) to the CFRS data in the size range $3.2 \leq \alpha^{-1} \leq 32\ h_{50}^{-1}$ kpc. The figures in brackets refer to the total number of galaxies used in the construction of the size function and the number in the $3.2 \leq \alpha^{-1} \leq 32\ h_{50}^{-1}$ kpc

Figure 4. The results of the fitting of a double power-law size function of fixed shape but varying characteristic size and comoving density to the $q_0 = 0.5$ size functions between $3.2 \leq \alpha^{-1} \leq 32\ h_{50}^{-1}$ kpc as shown in Fig. 3. The hatched areas represent, in order of decreasing shade for the increasing redshift ranges, an estimated $1\sigma$ confidence bound.

Figure 5  The change in characteristic size (at fixed comoving density) and in comoving number density (at fixed characteristic size) for the fits to the size functions of Fig 3. Open and filled symbols represent the effect of adding galaxies with only estimated redshifts (as in Fig 3). In the upper graph. formal best fits and 1s variations are shown for all points and for the CFRS points at $z > 0.2$. Especially if the $z = 0$ point from the de Jong (1996b) size function is included, there is little evidence for any systematic change in the size function to high redshift.

Figure 6  The distribution of axial ratios exhibited by the disk components in the sample of large disks ($\alpha^{-1} \geq 4\ h_{50}^{-1}$ kpc) in disk-dominated ($B/T \leq 0.5$) CFRS and LDSS galaxies, in three redshift bins. Two-dimensional disks randomly oriented in space should produce a uniform distribution in *b/a*, although intrinsic asymmetries and finite disk thickness and/or warping will cause systems to avoid *b/a* =1 and *b/a* = 0 respectively.

Figure 7  The variation of observed central surface brightness of CFRS and LDSS galaxies with inclination, i.e. disk axial ratio *b/a*. There is no significant correlation between these at any redshift, indicating that the disks are unlikely to be optically thin. The thin dashed line indicates the mean observed surface brightness. Solid symbols represent galaxies for which the two-dimensional fits had small residuals ($R_{A+S} \leq 0.1$) and open symbols represent galaxies with larger residuals. Galaxies classified by eye as spirals are represented by circles (CFRS) or pentagons

range. The three curves represent the best fitting curve allowing the size, the density, and both the size and density to vary.

(LDSS), those classified as "irregular/peculiar" by triangles (CFRS) and squares (LDSS).

Figure 8: The distribution of bulge-to-total *B/T* ratios in the sample of large disks. The three left hand panels show the distributions of *B/T* derived from the HST F814W images. The three corresponding panels on the right show the effect of applying a small correction to the rest-frame *B*-band (see text for details). The three distributions are consistent with each other and with the distribution observed in the local de Jong (1996a,b) sample shown in the uppermost panel.

Figure 9: The central surface brightnesses of the large disks ($\alpha^{-1} \geq 4\ h_{50}^{-1}$ kpc) in disk dominated ($B/T \leq 0.5$) CFRS and LDSS galaxies as a function of redshift compared with the low redshift de Jong sample (for which the redshift scale has been expanded for clarity). Symbols are as in Fig. 7: galaxies classified by eye as spirals are represented by circles (CFRS) or pentagons (LDSS), those classified as "irregular/peculiar" by triangles (CFRS) and squares (LDSS). The three dotted lines show the loci of three CFRS galaxies with unknown redshifts (LDSS galaxies without redshifts are not shown), while the two heavy dashed lines show the possible effects of surface brightness biases. The short dashed curve represents the minimum surface brightness of a pure disk galaxies with $\alpha^{-1} = 4\ h_{50}^{-1}$ kpc that satisfies the $I_{AB}$(total) $\leq 22.5$ selection criterion of the CFRS. The long dashed curve represents the surface brightness corresponding to the $1\sigma$ (per 0.01 arcsec$^2$ pixel) surface brightness limit of the F814W HST images. A significant brightening of surface brightness at higher redshifts is seen, irrespective of which particular set of galaxies are included, although some of the increase in the average may be due to the exclusion of low surface brightness galaxies as discussed in the text.

Figure 10: As in Fig. 9 except the integrated rest-frame (U-V)$_{AB}$ color is plotted. Within the high redshift sample there is a systematic bluening of colors at higher redshifts.

|            |                                                                                                                                                                                                                                                                                                                       |
|------------|---------------------------------------------------------------------------------------------------------------------------------------------------------------------------------------------------------------------------------------------------------------------------------------------------------------------|
|            | The fraction of galaxies lying bluewards of the $(U-V)_{AB} = 1.38$ criterion used to divide the CFRS luminosity function of Lilly et al (1995) also increases at high redshifts. |
| Figure 11: | The distribution of rest-frame equivalent widths of [OII] 3727 in the sample of large disk-dominated galaxies. The CFRS sample is shown as solid regions, the LDSS sample as hatched regions. Galaxies without redshifts are omitted (obviously). The median equivalent widths of the different subsamples increases modestly from low to high redshift. |
| Figure 12: | The distribution of eyeball morphological classifications for the large disk-dominated galaxies, represented as in Figure 11. There is a systematic trend to later morphological types at high redshift. |
| Figure 13: | Population synthesis models for evolving stellar populations. The relative surface brightness and colors of stellar population models with different exponentially declining star formation rates are shown as solid lines labeled with the time constant $\tau$ in Gyr. The thin dotted lines show the surface brightnesses and colors of the models at intervals of 2 Gyr, starting at 1 Gyr and ending at 13 Gyr and labeled with the age in Gyr. The models, computed from the GISSEL library, are for solar metallicity and have been reddened by an extinction of $E(U-V) = 0.2$ and have also had a red $(U-V)_{AB} = 2.5$ component contributing 20% of the light at V added to represent an average bulge contribution. The surface brightness of both the models and the data have been normalized to the nominal Freeman (1970) value at the present epoch (assumed to be 13 Gyr). The data points show the average properties of the CFRS + LDSS galaxies from Table 2, differentiating between all galaxies and those with good residuals and spiral morphology (Table 2), but omitting the latter at the highest redshift (since there are only 3 galaxies in this category). It should be remembered that the surface brightness of the point at $z = 0.87$ may be overestimated by as much as 0.4 mag and that at $z = 0.63$ by 0.2 |

mag. The mean properties of the large disks galaxies are *broadly* consistent with models with exponential star-formation decay timescales of 5 Gyr and quite inconsistent with a constant star-formation rate.

Figure 14: The half-light radii, computed from the surface brightness fits, for *all* the CFRS galaxies in the sample as a function of their *B*-band absolute magnitude in panels differentiated by redshift and by rest-frame $(U-V)_{AB}$ color. The galaxies represented by asterisks are those in the large disk sample studied in this paper, which are believed to be equivalent to the Vogt et al (1996, 1997) sample of disks studied kinematically, while the triangles represent those that should have appeared in the Phillips et al (1997) and Guzman et al (1997) samples of "compact" galaxies. Open symbols represent galaxies with estimated redshifts. The heavy dashed line shows roughly the luminosity limit that corresponding to the CFRS $I_{AB} \leq 22.5$ apparent magnitude limit. The distribution of points in these diagrams suggests that the main evolutionary changes to the galactic population over the $0 \leq z \leq 1$ interval concern galaxies with half-light radii smaller than 5 $h_{50}^{-1}$ kpc rather than the large disks that are the focus of this paper.

Figure 15: The bivariate size-luminosity function $\phi(M_B, r_{0.5})$ derived for the CFRS subsample (as in Table 3). For simplicity, objects without secure redshifts are included at their estimated redshifts. While the "ridge-line" at large sizes remains roughly constant (increasing by around 0.2 dex in height), the biggest evolutionary change is the filling in of the "reentrant" at $M_{AB}(B) \sim -21.5$ and $r_{0.5} \sim 3$ kpc with substantial numbers of galaxies at high redshift. This is especially evident in the $q_0 = 0.5$ case, and the impression of differential behavior in the size-luminosity plane is significantly weaker for $q_0 = 0.0$.

**Table 2: Average properties of the large disk sample**

| Redshift | Sample | Quality[a] | $N_{gal}$ | $\mu_{0,AB}(B)$[b] | $(U-V)_{AB}$[b] | $EW_{OII}$[c] |
|---|---|---|---|---|---|---|
| $z = 0$ | de Jong 1996 | ... | 48 | $21.78 \pm 0.88$ | $1.54 \pm 0.41$ | ... |
| $0.2 < z < 0.5$ | CFRS | All | 11 | $21.48 \pm 0.88$ | $2.15 \pm 0.44$ | 9 |
| | CFRS+LDSS | All | 15 | $21.30 \pm 0.89$ | $1.90 \pm 0.57$ | 15 |
| | CFRS | Best | 7 | $21.71 \pm 0.91$ | $2.28 \pm 0.32$ | <5 |
| | CFRS+LDSS | Best | 11 | $21.38 \pm 0.95$ | $1.88 \pm 0.57$ | 12 |
| $0.5 < z < 0.75$ | CFRS | All | 16 | $20.81 \pm 0.85$ | $1.36 \pm 0.48$ | 15 |
| | CFRS+LDSS | All | 19 | $20.64 \pm 0.95$ | $1.30 \pm 0.47$ | 15 |
| | CFRS | Best | 10 | $20.85 \pm 0.64$ | $1.40 \pm 0.44$ | 13 |
| | CFRS+LDSS | Best | 11 | $20.67 \pm 0.83$ | $1.33 \pm 0.47$ | 13 |
| $0.75 < z < 1.0$ | CFRS | All | 8 | $20.24 \pm 0.45$ | $1.25 \pm 0.76$ | 19 |
| | CFRS | Best | 3 | $20.64 \pm 0.25$ | $2.04 \pm 0.64$ | 12 |

Notes to Table 2

[a]"All" means all objects, "best" means only those that RA+S ≤ 0.10 and morphological classification as Sdm or earlier type.

[b]Mean of galaxies in sample

[c]Median of galaxies in sample

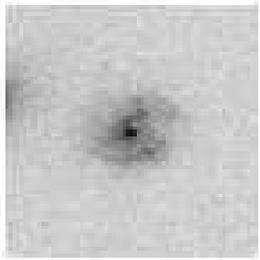 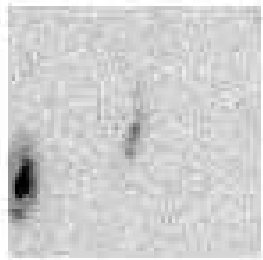 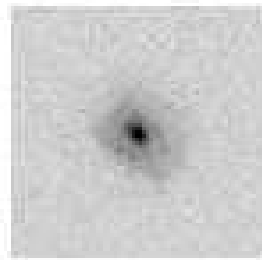 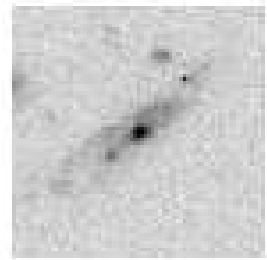
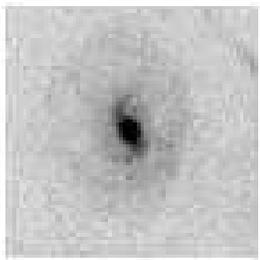 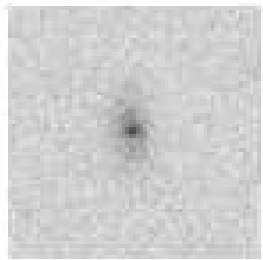 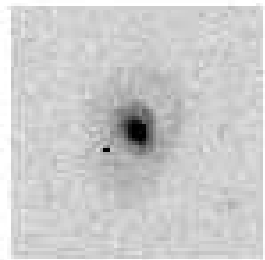 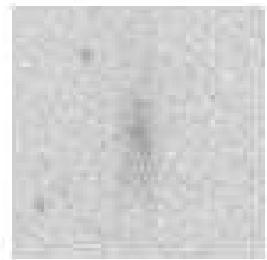
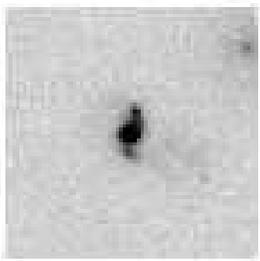 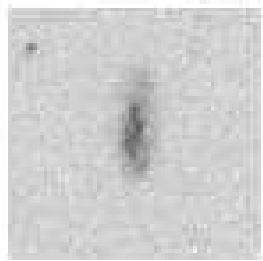 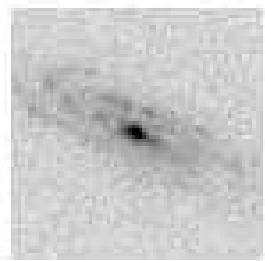 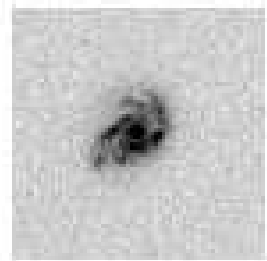
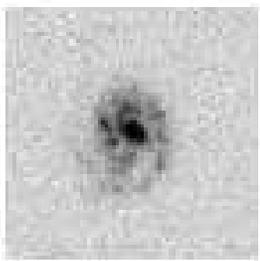 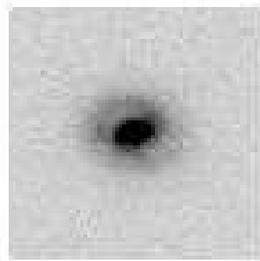 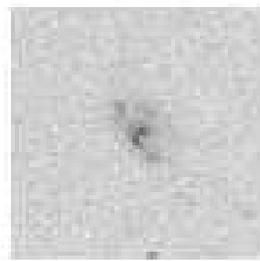 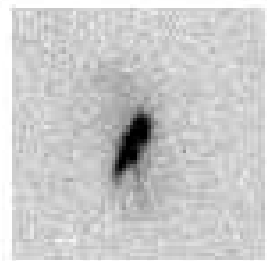
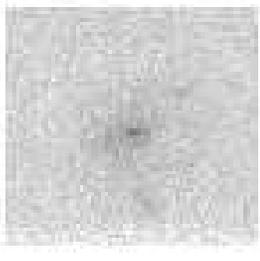 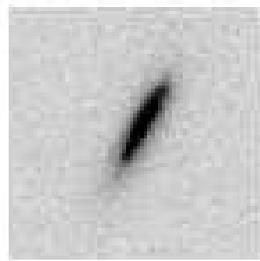 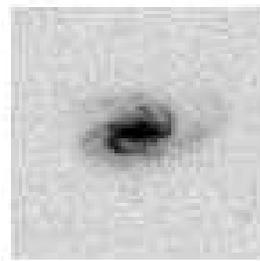

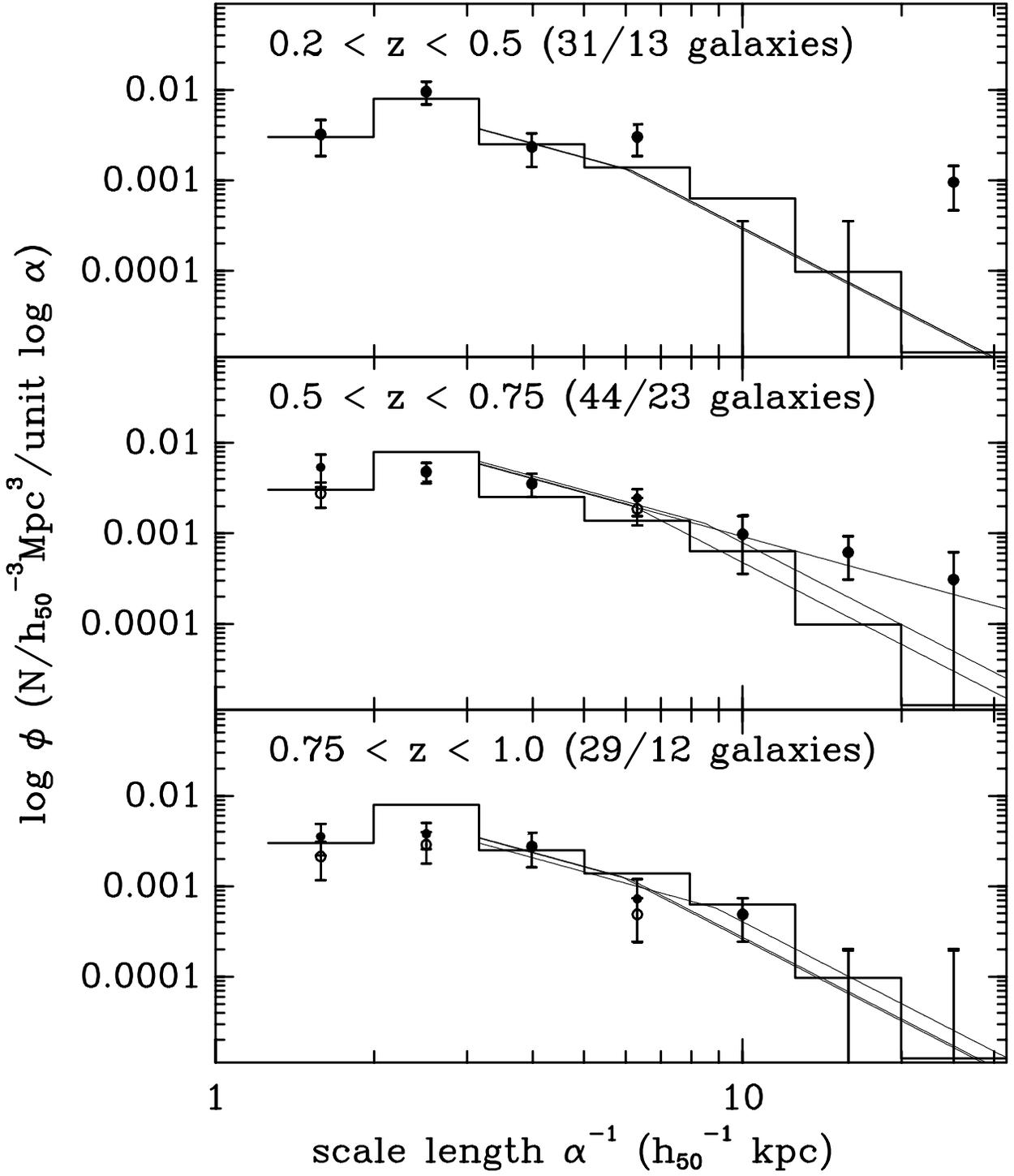

Lilly et al Fig 3

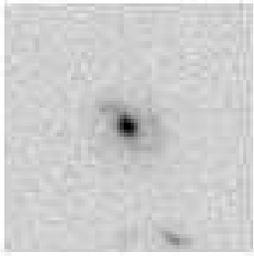 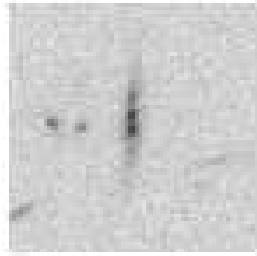 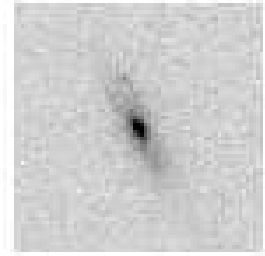 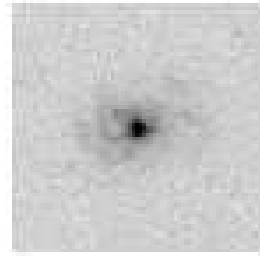
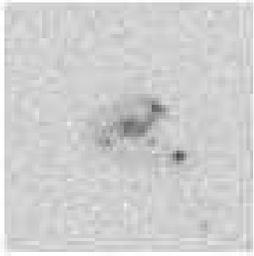 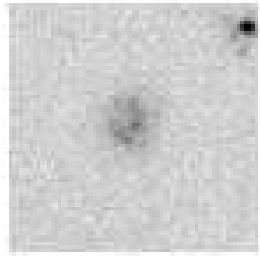 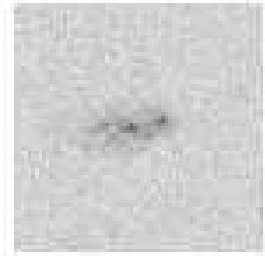 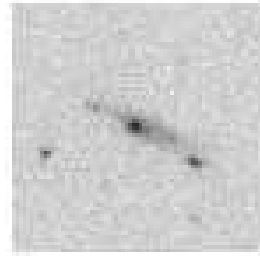

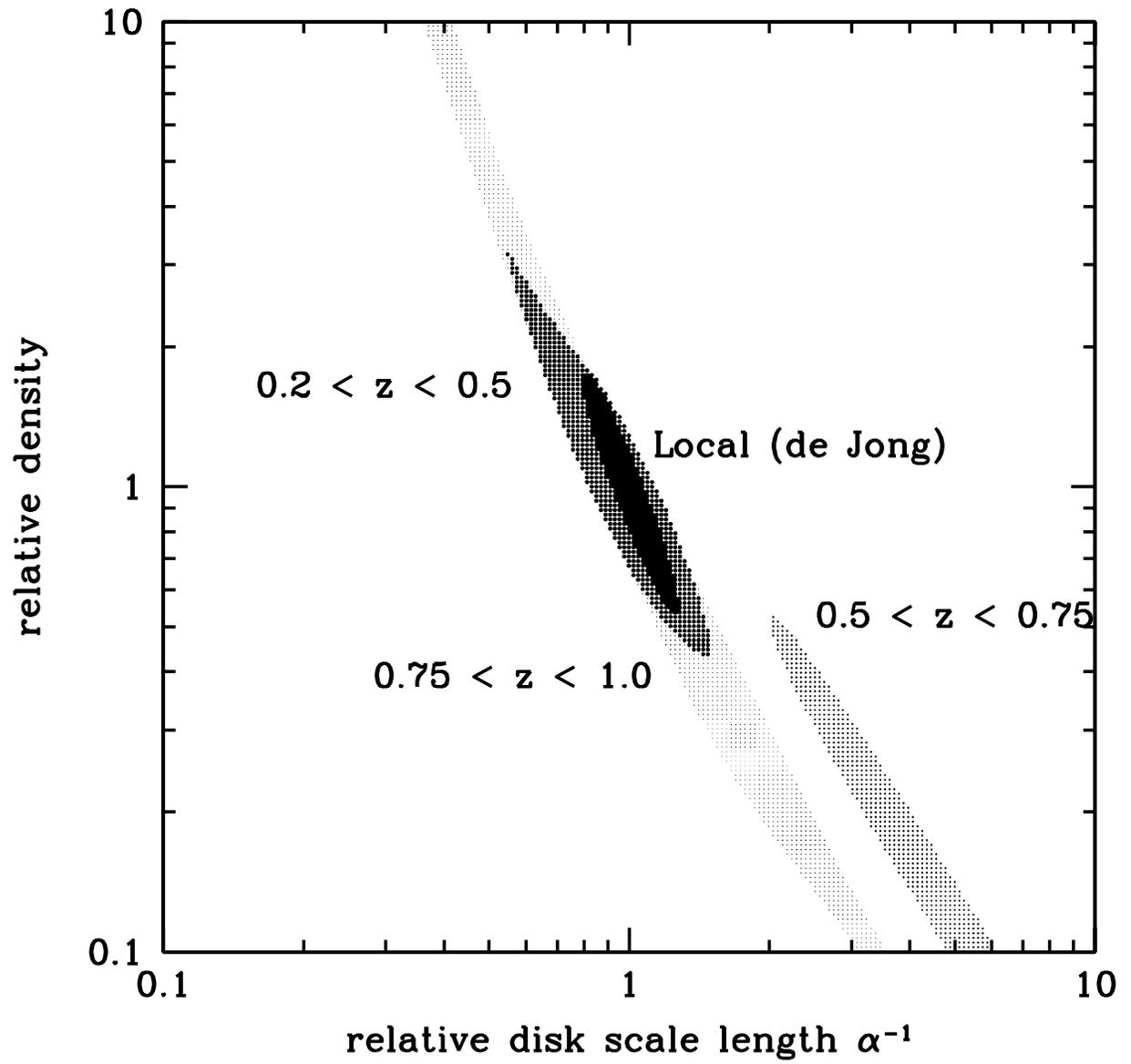

Lilly et al Fig 3

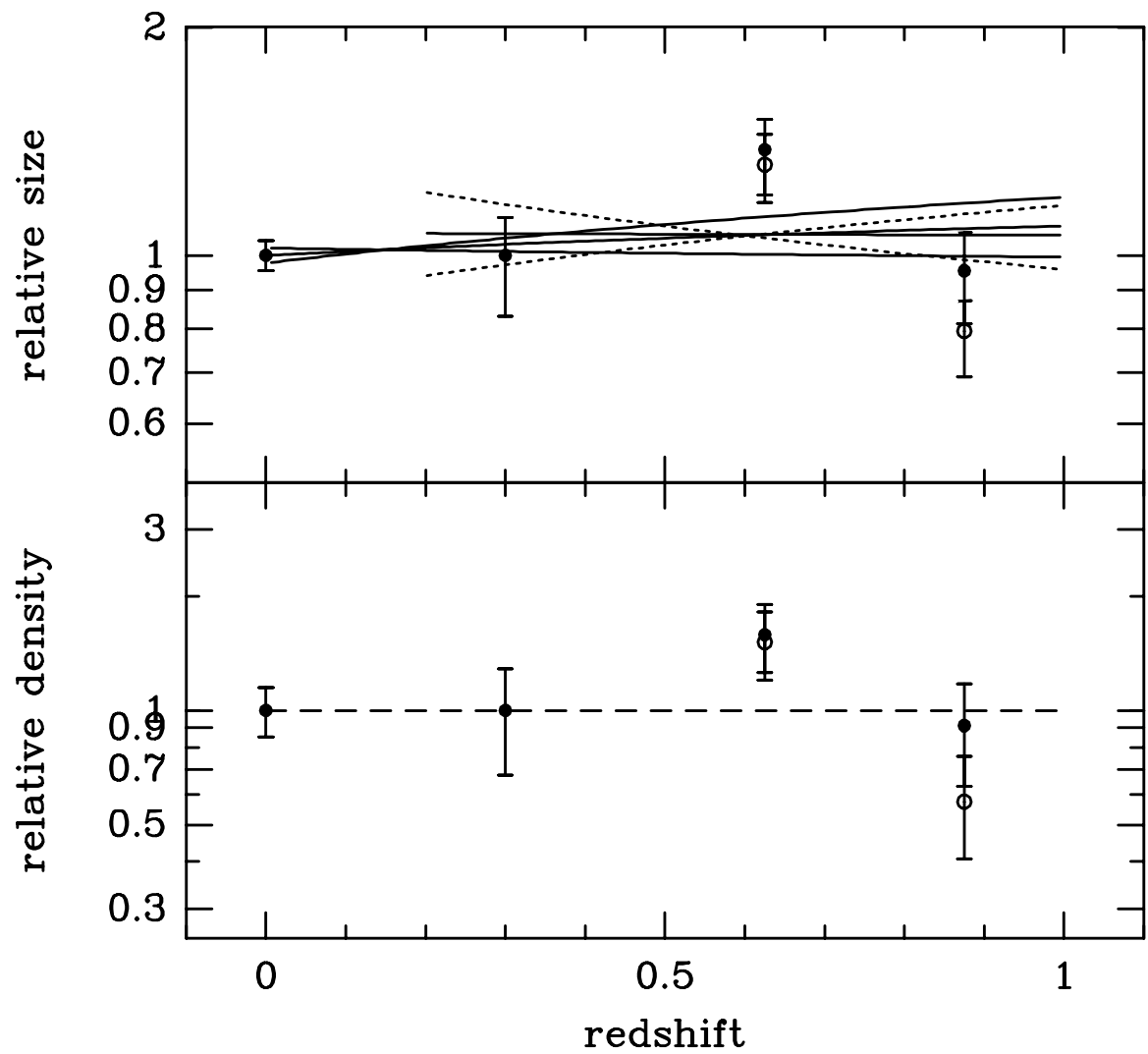

Lilly et al Fig 5

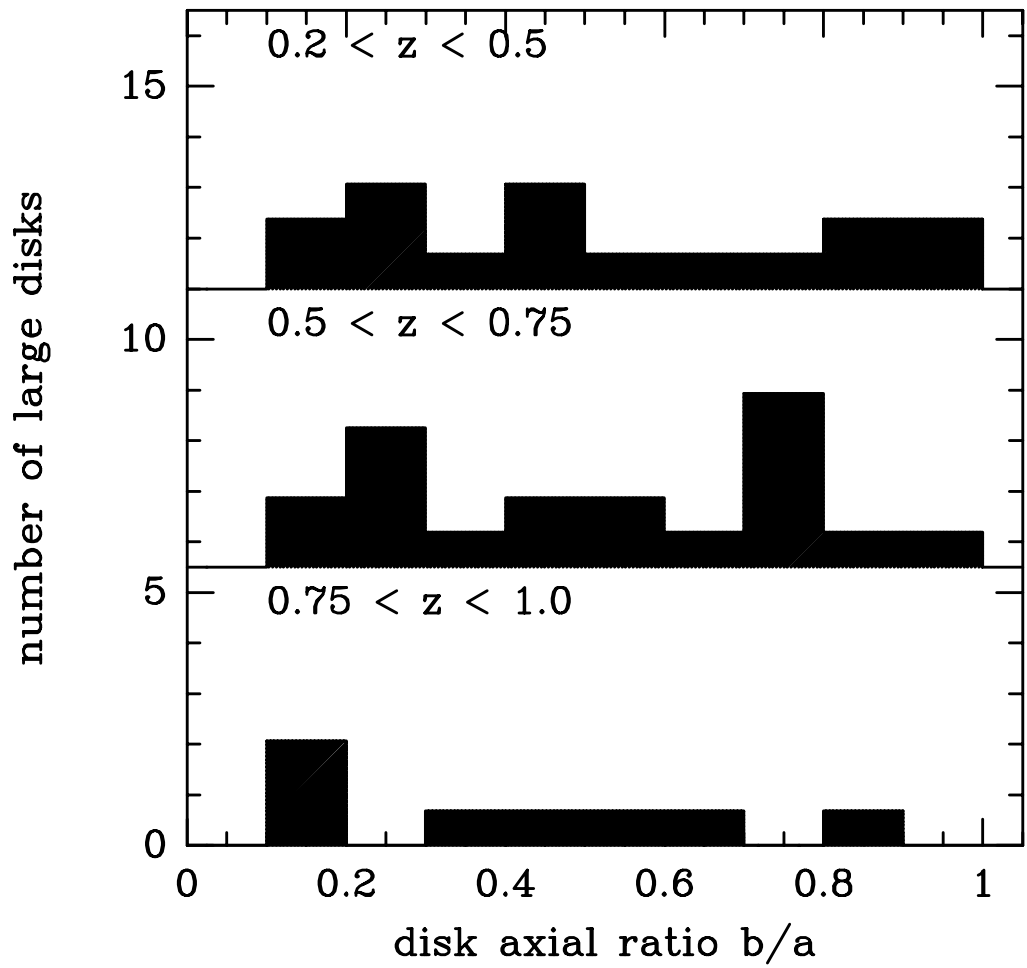

Lilly et al Fig 6

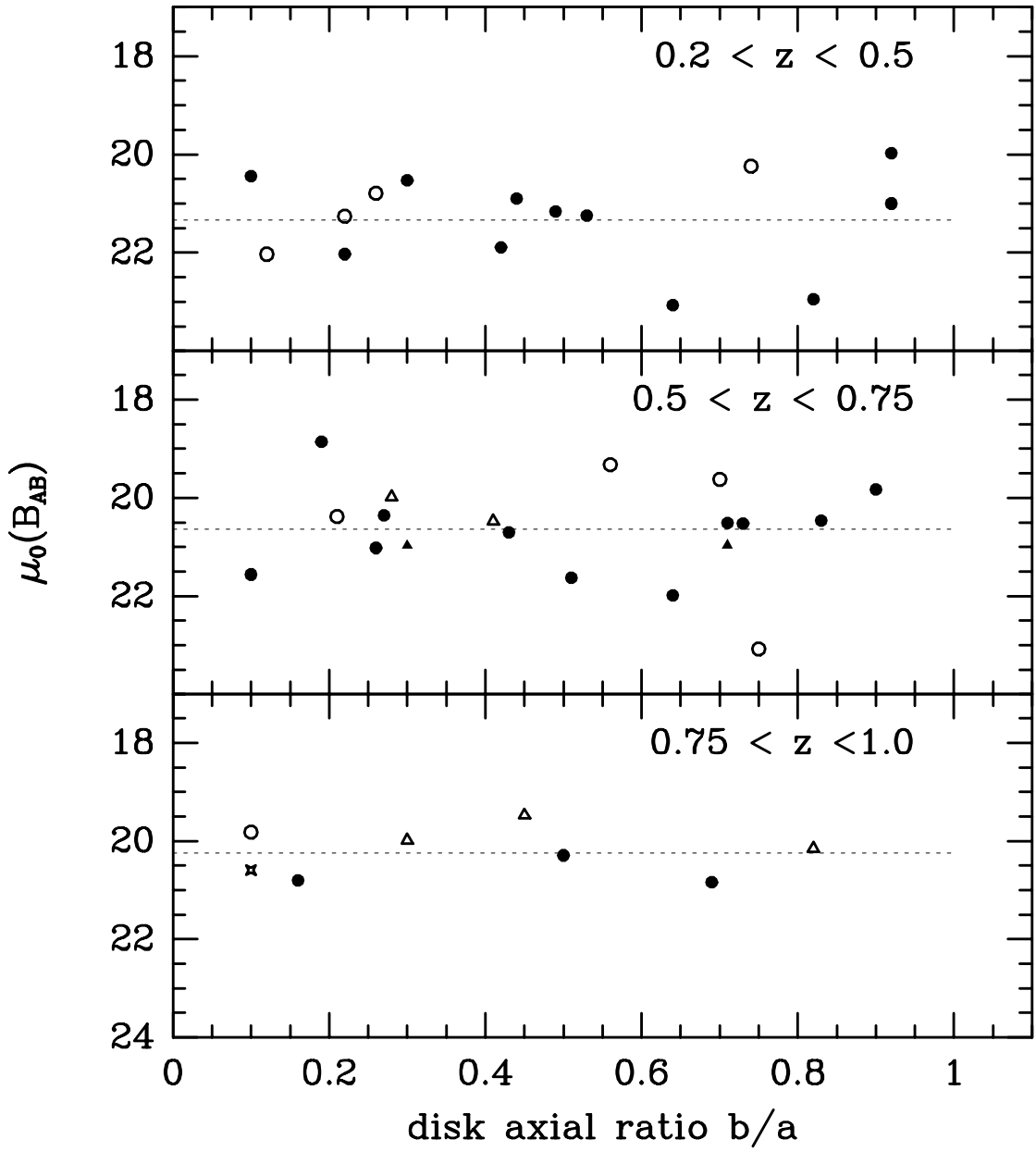

Lilly et al Fig 7

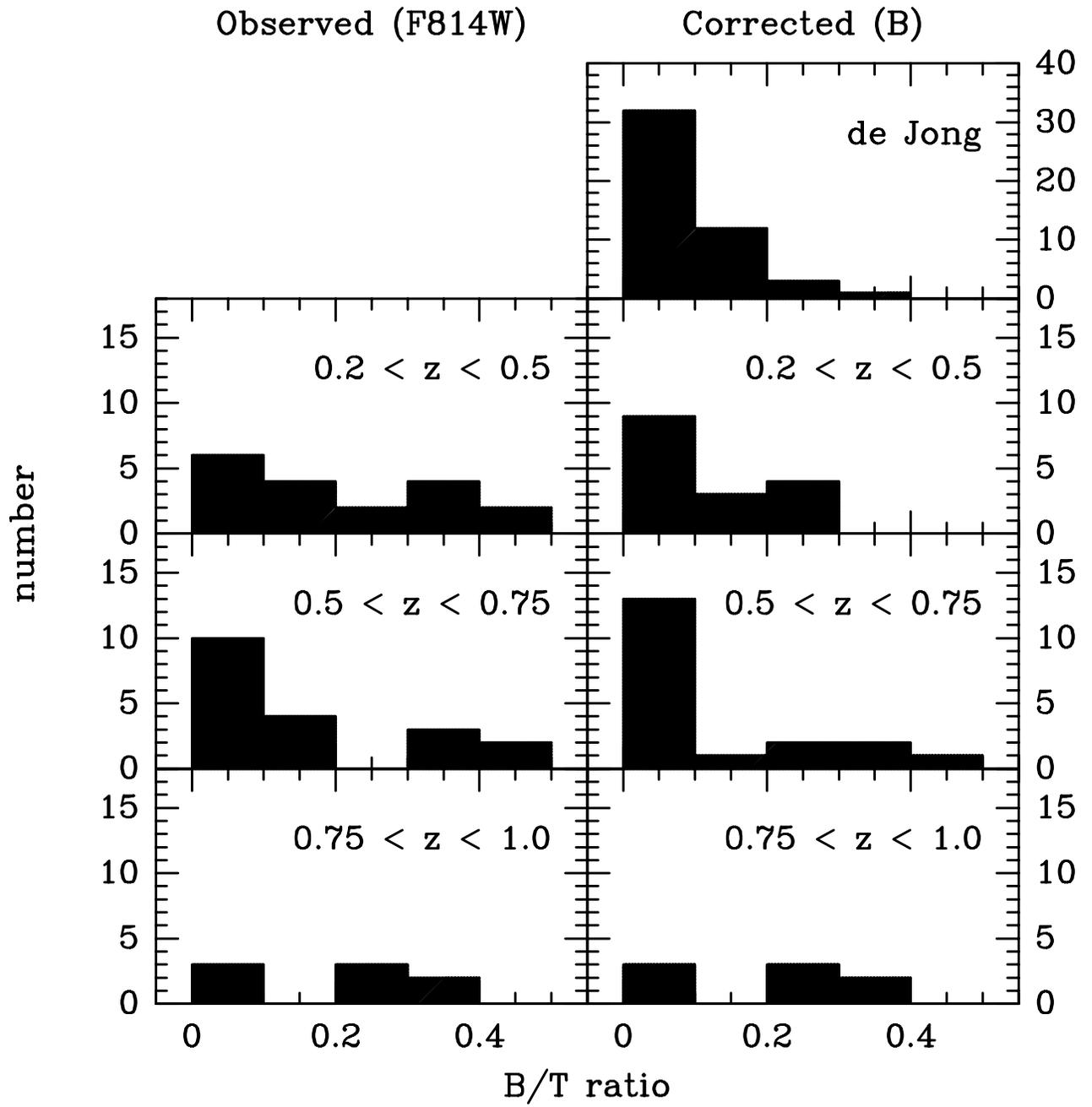

Lilly et al Fig 8

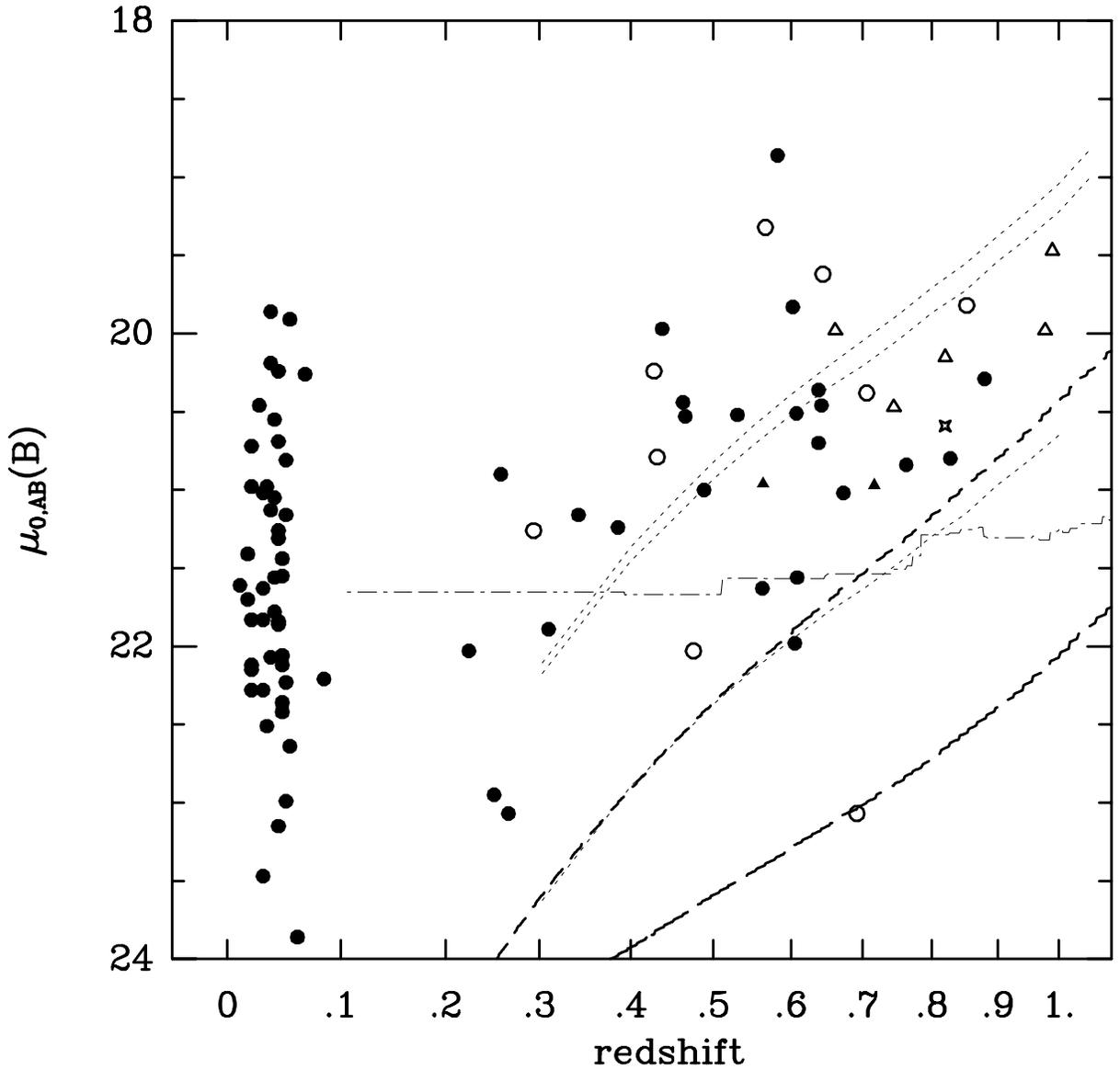

Lilly et al Fig 9

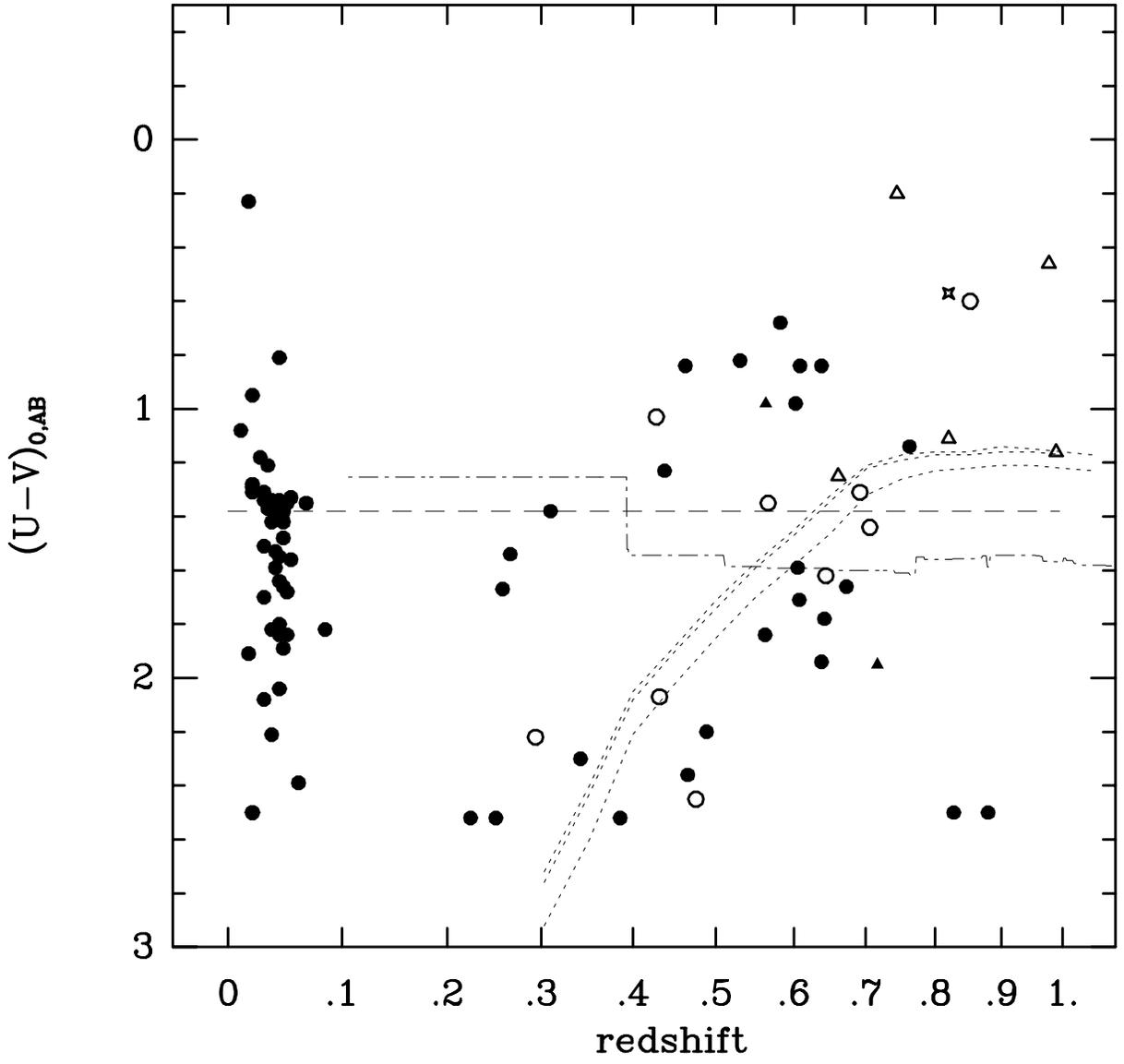

Lilly et al Fig 10

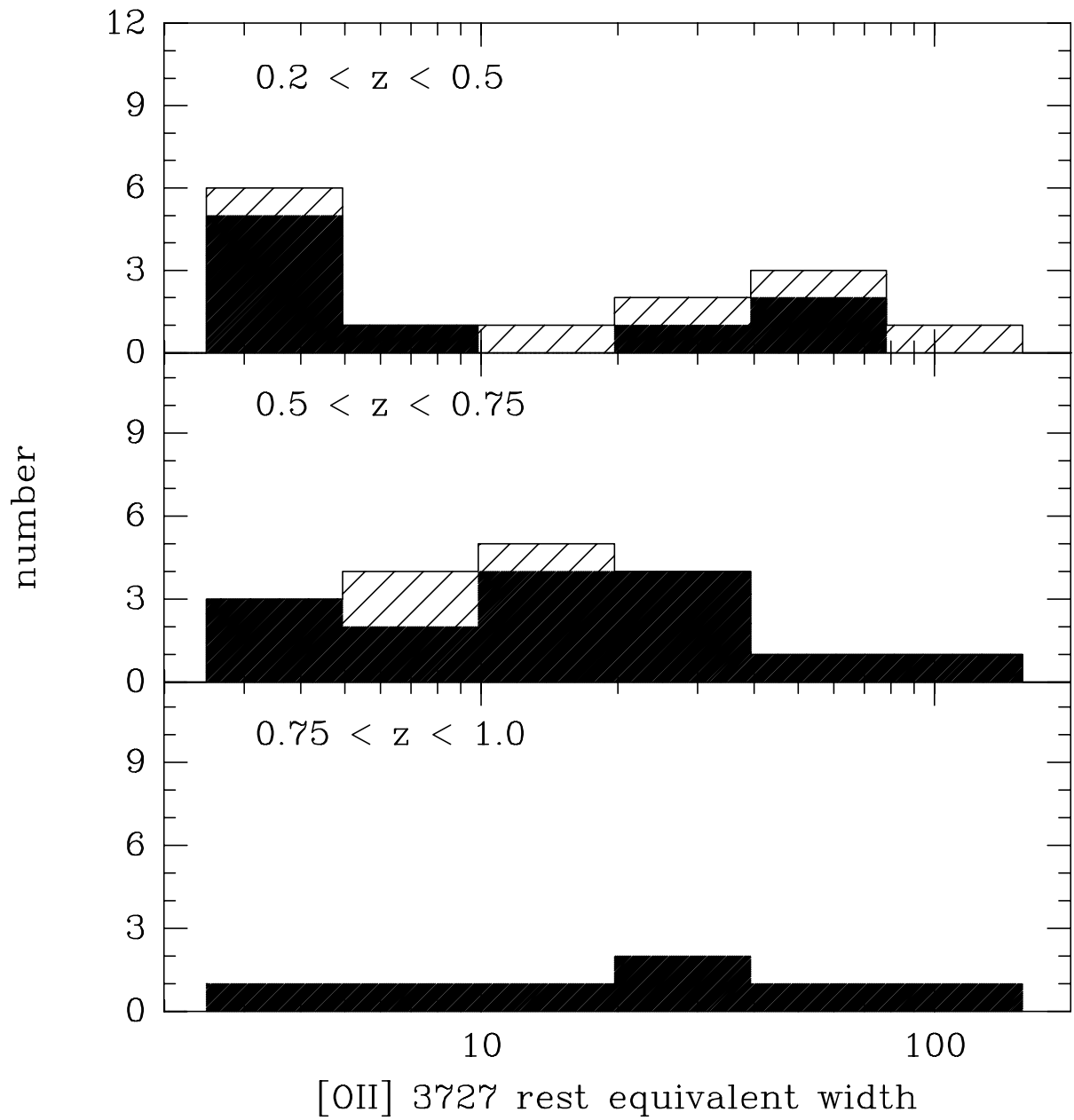

Lilly et al Fig 11

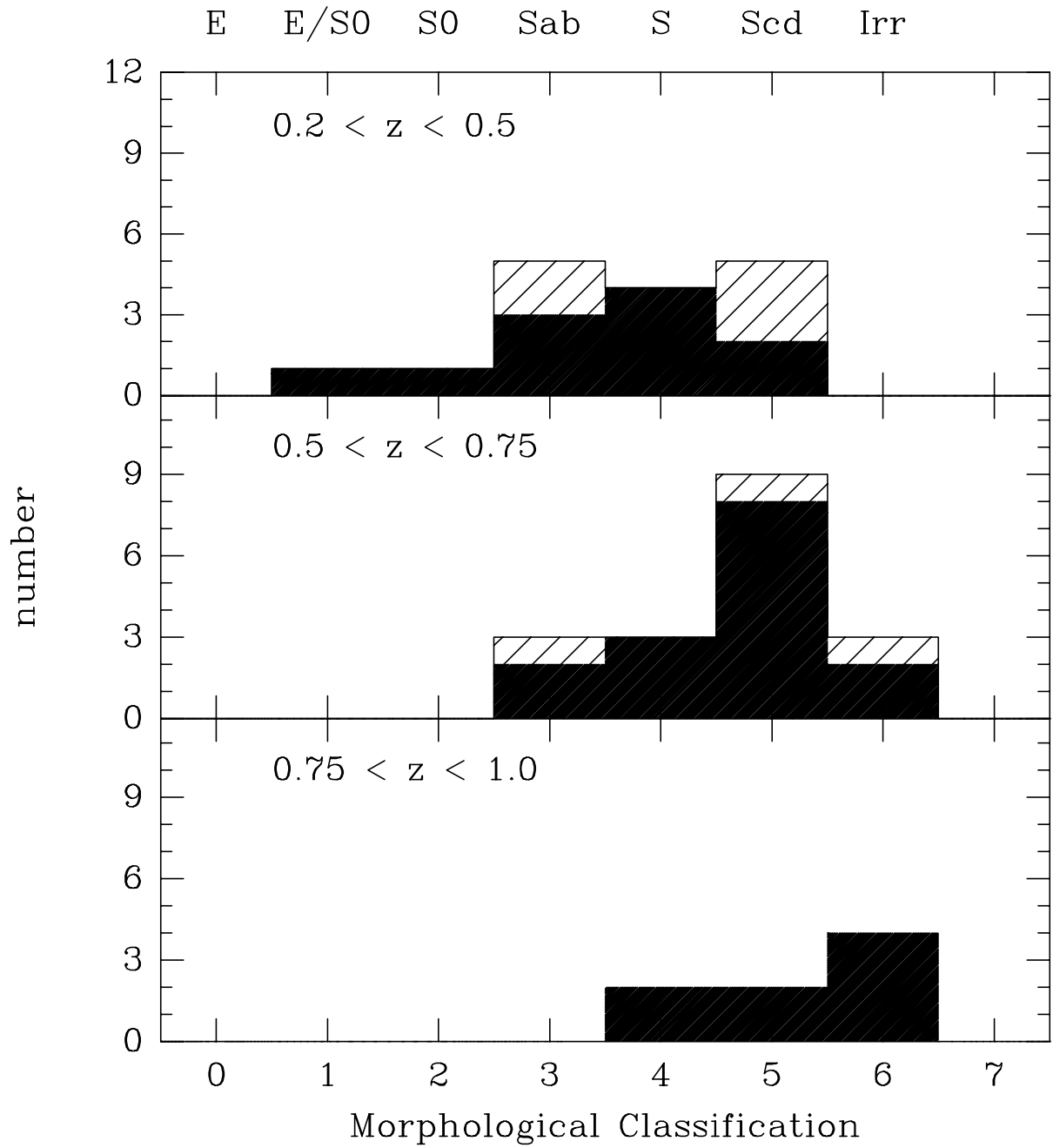

Lilly et al Fig 12

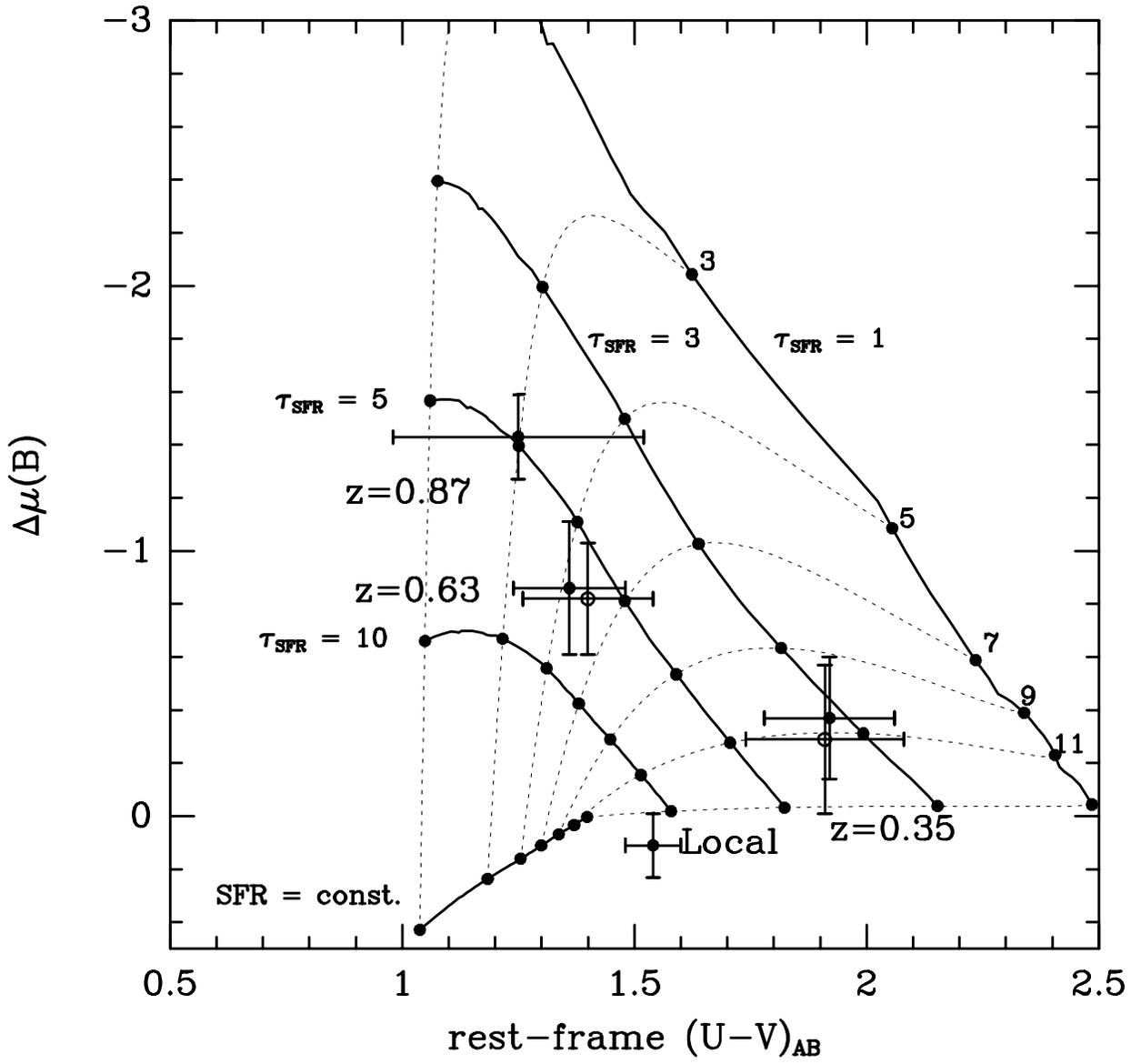

Lilly et al Fig 13

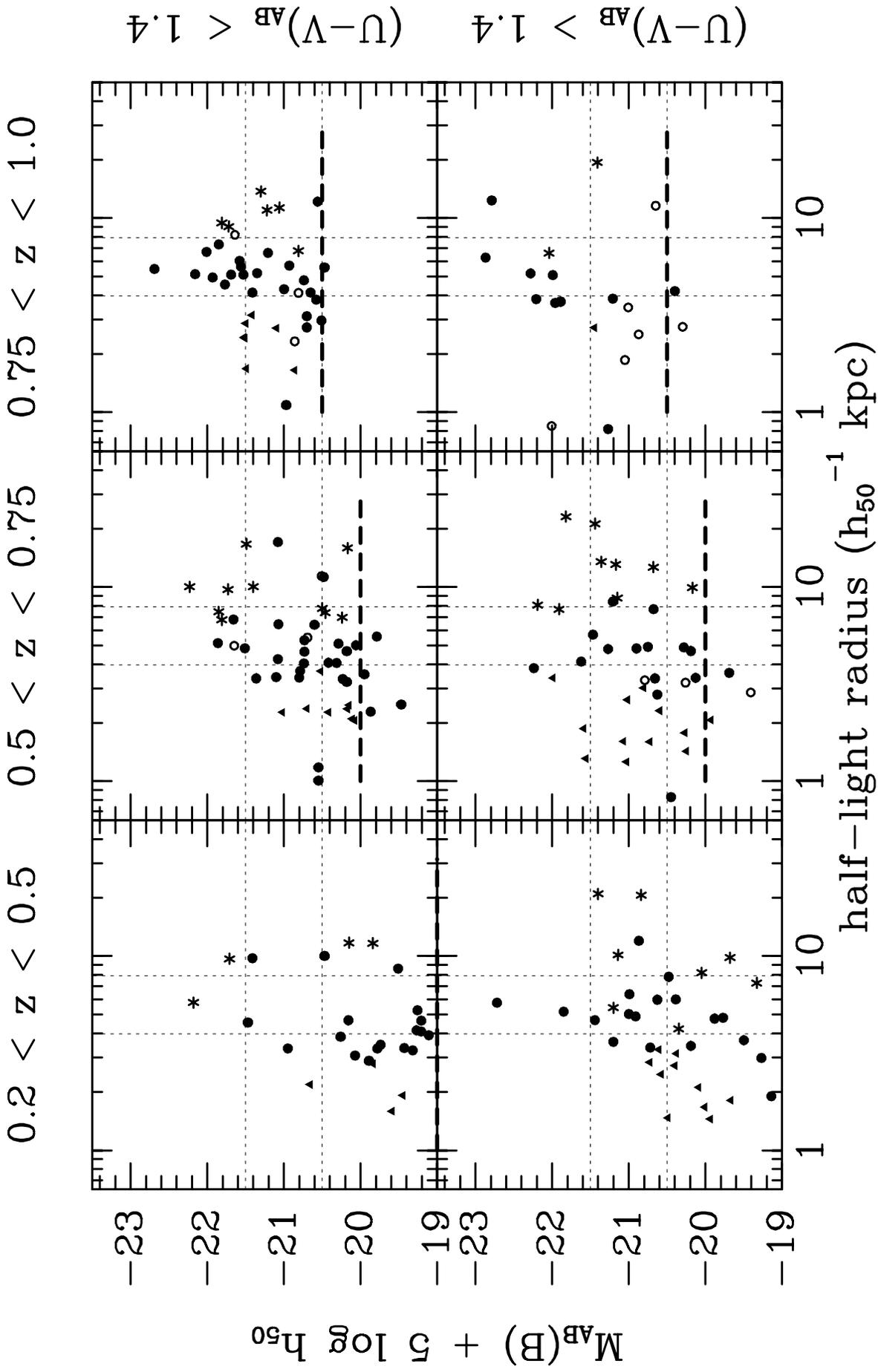

Lilly et al Fig 14

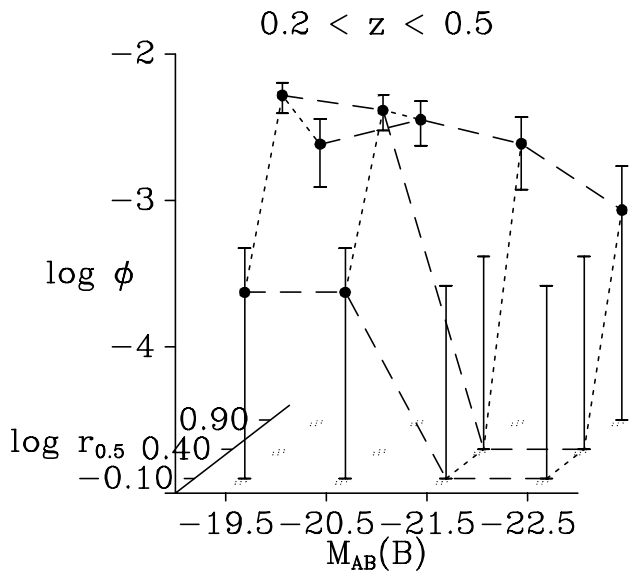
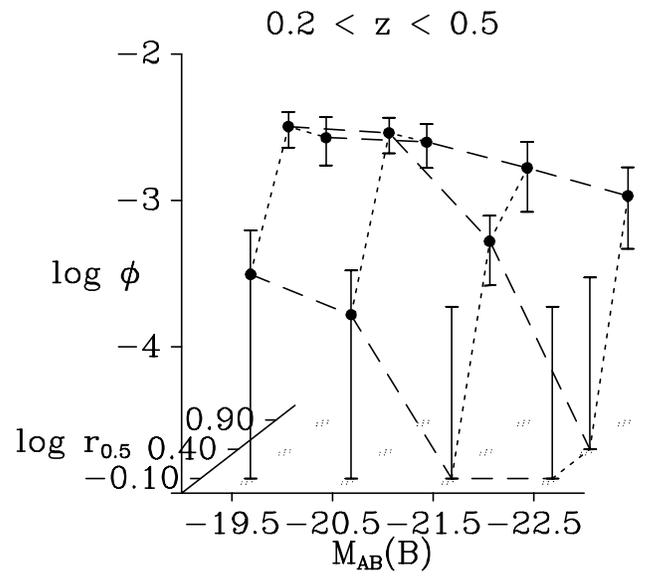
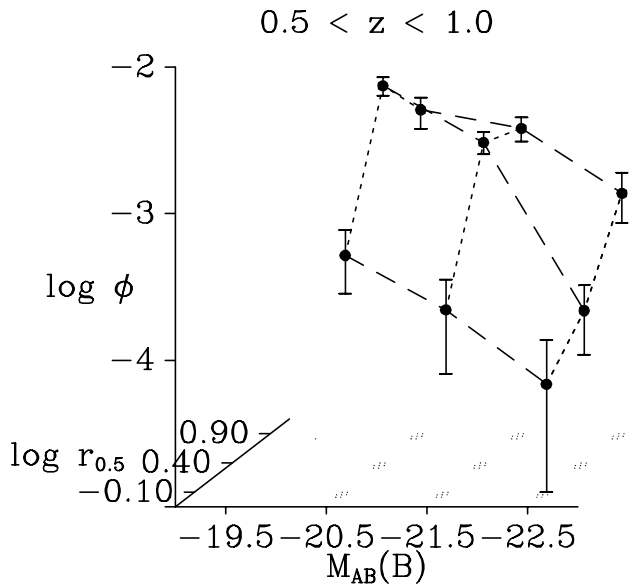
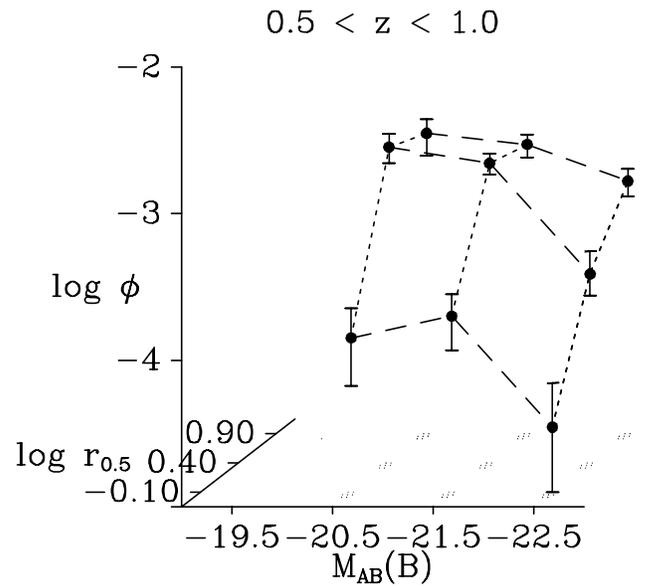

Lilly et al Fig 15